\documentclass[aps,prb,reprint,groupedaddress,longbibliography]{revtex4-1}

\usepackage{amsmath,amssymb,slashed}
\usepackage{color}
\usepackage{graphicx}
\usepackage{latexsym, bm} 

\begin{document}

\title{Quantum Interactions of Topological Solitons from Electrodynamics}


\author{
Hirohiko Shimada$^1$,
Kazutaka Takahashi$^2$,
Hiroaki T. Ueda$^{3}$
}\thanks{Deceased 11 December 2016.}

\affiliation{
$^1$Mathematical and Theoretical Physics Unit, OIST Graduate University, Onna, Okinawa 904-0495, Japan\\
$^2$Department of Physics, Tokyo Institute of Technology, Tokyo 152-8551, Japan\\
$^3$Faculty of Engineering, Toyama Prefectural University, Izumi 939-0398, Japan}


\date{\today}

\begin{abstract}
The Casimir energy for the classically stable configurations of the topological solitons in 2D quantum antiferromagnets 
is studied by performing the path-integral over quantum fluctuations. 
The magnon fluctuation around the solitons saturating the Bogomol'nyi inequality may be viewed as     
a charged scalar field coupled with an effective magnetic field induced by the solitons. 
The magnon-soliton couping is closely related to the Pauli Hamiltonian, with which 
the effective action is calculated by adapting the worldline formulation of the derivative expansion 
for the 2+1 dimensional quantum electrodynamics in an external field.
The resulting framework is more flexible than
the conventional scattering analysis based on the Dashen-Hasslacher-Neveu formula.
We obtain a short-range attractive well and a universal long-range $1/r$-type repulsive potential between two solitons.
\end{abstract}

\pacs{}

\maketitle

\section{Introduction}

Topological solitons in magnets, also known as magnetic skyrmions or lumps, are extended particle-like excitations
whose stability may be protected by non-trivial topology.
In quantum field theory (QFT),   
they were first found as non-trivial classical field configurations \cite{belavin1975metastable} 
that minimize the action 
in the continuum limit of the 2d $O(3)$ Heisenberg spin model. 
In this example, the action is bounded below by the Bogomol'nyi inequality, and has the minima classified by the topological 
charge $q\in \pi_2(S^2)=\mathbb{Z}$. Each minimum is realized by uncountablly many energetically equivalent configurations of solitons.  
Such a large degeneracy of configurations is exact at the classical level, and is generalized to 
the idea of the moduli space of solitons, whose rich mathematical structure and 
geodesic approximations to the dynamics on it have been extensively studied \cite{ward1985slowly, manton2004topological,weinberg2012classical}.

More recently, stimulated by the experimental observations of the skyrmions in chiral magnets \cite{muhlbauer2009skyrmion},   
the notion of the emergent electromagnetism \cite{schulz2012emergent,nagaosa2013topological,volovik1987linear} becomes 
increasingly important, envisaging potential applications of the solitons for memory and logic devices \cite{fert2013skyrmions}.
The emergent electromagnetism in antiferromagnetic solitons may also attract an attention as a new direction in spintronics \cite{barker2016static, zhang2016antiferromagnetic, jin2016dynamics, dasgupta2017gauge, smejkal2018topological}.

The effect of the quantum fluctuations along with the emergent electromagnetism 
on the stability of the soliton configurations,
 however, remains relatively unexplored and somewhat elusive.
For instance, most of the current 
studies of the quantum effects hinge on 
single old strategy \cite{dashen1974nonperturbative, rajaraman1982solitons} 
used initially by Dashen-Hasslacher-Neveu (DHN) in the semiclassical analysis of 1+1d QFTs. 
Generalizations of the DHN formula to 2+1d may allow one to evaluate the 
energy shifts (Casimir energy) due to the zero-point oscillations of 
the vacuum magnon states in the presence of the solitons
 from the scattering data, namely, the phase shifts of the spin waves around the solitons, for which one typically needs to invoke the Born approximation \cite{rodriguez1989quantized}, the Aharonov--Bohm (AB) scattering scenario \cite{ivanov2007quantum}, 
 or a heavy use of numerics \cite{walliser2000casimir}. 

Theoretical outcomes also seem far from being settled in general; for example in isotropic ferromagnets, 
a recent intriguing argument emphasizing the role of 
the Bogomol'nyi equality
points to the absence of the quantum corrections 
 \cite{douccot2016large}, which is in an apparent contradiction to the finite Casimir energy obtained in Ref. \onlinecite{ivanov2007quantum} also predicting a spontaneous collapse of the tiniest-possible soliton 
triggering a quantum increment or reduction of the topological charge of the system.

In this paper, we study the quantum effect on static soliton configurations, which are classically degenerate, in 2d isotropic antiferromagnets. 
Our path-integral approach is based on a geometric observation that the quantum fluctuations 
around the soliton configuration saturating the Bogomol'nyi inequality (BPS solitons) resembles those in quantum electrodynamics (QED) in an external field;
the magnon excitation 
may be described by a complex scalar field coupled with 
an effective abelian gauge field as well as with the magnetic field proportional to the topological charge density of the solitons.
It is then natural to expect that, upon integrating the magnon degrees of freedom out, this coupling should yield the effective action describing the quantum interactions of the solitons with themselves
in the same way as the Maxwell Lagrangian acquires nonlinear quantum corrections representing the light-by-light scattering in the Euler-Heisenberg Lagrangian in QED \cite{heisenberg2006consequences,nambu1950use,schwinger1951gauge}.

Such nonlinear terms
arise from multiple pair-productions of 
scalar \cite{nambu1950use} or spinor \cite{schwinger1951gauge}
particles, which are inherently non-perturbative,  
but are well-controlled (with no small parameters) in the case of the uniform external field.
Slow variations of the soliton magnetic field $B$ can then be taken into account by
adapting the derivative expansions in 2+1d QED \cite{cangemi1995derivative, gusynin1999derivative}.
This is done by the worldline (proper-time) path-integral formalism 
\cite{strassler1992field, schmidt1993calculation, schubert2001perturbative}
by incorporating inhomogeneities as perturbations
on the particle motion, solved for a constant external field, along the one-loop Feynman diagram.

At the heart of this formulation, we have a certain equivalence between a one-loop sector of a relativistic QFT in $d+1$-dimensions ($d=2$ in our case) and a non-relativistic quantum mechanics (or $0+1$-dimensional QFT) of $d+1$ degrees of freedoms living on the loop parameterized by the proper-time. 
In contrast to the more standard perturbation theory in which one performs the expansion 
in powers of the coupling constants such as $B$ (charge $e$ is incorporated in $B$ for now), it is based on the semi-classical expansions in powers of $(\hbar/S)$ associated with each loop, where $S$ is the spin
\footnote{The ratio $(\hbar/S)$ arises naturally since the action $\mathcal{S}$ is proportional to spin $S$. It is also well-known\cite{sachdev2007quantum} that the classical picture becomes better for large $S$ as the non-commutativity is suppressed in the fundamental commutation relation 
$[\hat{S}_i^a, \hat{S}_j^b]=i\hbar \delta_{ij}\epsilon^{abc}\hat{S}_j^c$ ($a, b, c \in \{ x,y,z\}$). }.
Even at the leading order, which we compute, this formulation adds up the one-loop diagrams to infinite order in the coupling $B$; as a result 
\footnote{Note that one can not obtain the fractional power $B^{3/2}$ at any finite order of the standard expansion in $B$.}
, one obtains the effective action $B^{3/2}$ in the limit of our interest, where the magnon mass goes to zero.
Technically, it also enables us to work out, otherwise difficult,
 the contribution proportional to $(\partial B)^2$ in the massless limit.
The existence of the coupling to the effective magnetic field, similar to what one has in the Pauli Hamiltonian,  
leads us to consider a one-parameter deformation of the scalar QED, and interestingly, the magnon-soliton system corresponds to the point exactly where the result simplifies significantly.

While our derivative expansion formula for antiferromagnets applies for 
any configurations of multi-solitons as long as they saturate the Bogomol'nyi inequality,
the simplest analytic result for one soliton suggests that it may shrink and eventually evaporate by quantum fluctuations; 
this also quantitatively agrees with the numerical work \cite{walliser2000casimir}
updated from the less precise Born approximation \cite{rodriguez1989quantized}.  
We obtain a long-ranged repulsive potential with an attractive well;
this is in qualitative agreement with the only existing work \cite{rodriguez1989quantized} for two solitons.

The paper is organized as follows. 
Roughly, Section \ref{section:magnon-soliton} is about the aspects
of the classical solitons, which are more or less known;
Section \ref{section:worldline} and \ref{section:quantum} 
are about their quantization, where the magnon and the solitons play the central role, respectively.
In Section \ref{section:magnon-soliton},
the action of the $O(3)$ model with the soliton background is expanded to the quadratic order in fluctuation,
yielding the magnon-soliton coupling \eqref{k_quadraticB}.
Using the equivalent $\mathrm{CP}^1$ representation, 
the various forms of the effective magnetic field are given, which become the key inputs
in Section \ref{section:quantum}. 
In Section \ref{section:worldline}, the derivative expansion \eqref{V=i}
of the quantum effective action for the soliton-magnon system
is derived by the proper-time integration along the magnon loop. 
In Section \ref{section:quantum}, we apply this formula to study the quantum interaction in
various soliton configurations.
We conclude in Section \ref{section:conclusion}.

\section{Magnon-soliton coupling} \label{section:magnon-soliton}
\subsection{2d quadratic Hamiltonian in the non-trivial topology}

We here focus on the spatial part of the full action for the $2+1$d anti-ferromagnet discussed in
Section \ref{section:quantumfluctuation}.
This amounts to consider 
the 2d classical $O(N)$ model, where the degrees of freedoms are the $N$-component
real vector $\bm{n}(x)$ 
with $|\bm{n}(x)|=1$.  
The energy is given by 
\begin{align}
H=& \frac{1}{4}\int {d}^2x~(\partial_{j}\bm{n})^2.
\label{H_O3}
\end{align}
 Below we follow the standard approach \cite{polyakov1975interaction} (see Ref. \onlinecite{polyakov1987gauge} for an insightful account) and decompose the variation of $\bm{n}(x)$ into the
 slow-longitudinal mode $\bm{n}_0(x)$
 and the fast-transverse-real-modes $\phi^{a}(x)$ living in the tangent space spanned by a local orthonormal frame
$\{\bm{e}^a\}$ with $\bm{e}^a\cdot \bm{e}^b=\delta^{ab}$ ($a=1, 2,\ldots, N-1$):
\begin{align}
\bm{n}(x)=\sqrt{1- \phi^a(x)\phi^a(x)}~ \bm{n}_0(x) + \phi^a(x)\bm{e}^a(x)
\label{slowfast}
\end{align}
By orthogonality $\bm{n}_0\cdot \bm{e}^a=0$, the most general form of the derivatives is 
\begin{align}
\partial_{j}\bm{n}_0=C^a_j\bm{e}^a,\qquad
\partial_{j}\bm{e}^a=-C^a_j\bm{n}_0 + A^{ab}_j\bm{e}^b,
\end{align}
 where the coefficient $C^a_j= \bm{e}^a \cdot\partial_{j}\bm{n}_0$
 only appears in the intermediate steps until the BPS condition is used,
 while a connection for the local frame
  $A^{ab}_j=\bm{e}^b\cdot\partial_{j}\bm{e}^a=-A^{ba}_j$
 play a role as a soliton-induced gauge field that couples to a scalar field. This scalar field is referred to as a magnon
described in Section \ref{section:quantumfluctuation}. 
Below we see that a dual pair of $C^a_j$ in \eqref{BPS0} constitutes the soliton magnetic field $F_{ij}$ in \eqref{nnn0} arising in the non-minimal coupling in \eqref{quadraticF}, where the field strength $F_{ij}$ directly couples to the magnon in addition to the gauge coupling through the connection $A_j$.

The derivative of \eqref{slowfast} is then
\begin{align}
\partial_{j}\bm{n}=&\left(\frac{\phi^a\partial_{j}\phi^a}{\sqrt{1-\phi^a\phi^a }}  
-C^{a}_j\phi^a  \right) \bm{n}_0 \nonumber\\
&+\left(\sqrt{1- \phi^c\phi^c}C^{a}_{j}  
+\partial_{j}\phi^a -A_{j}^{ab}\phi^b \right) \bm{e}^a, 
\end{align}
 resulting in the quadratic part of the energy \eqref{H_O3} that involves the 2d covariant derivative.
 The energy density in terms of the quadratic fluctuations is given by
\begin{align}
\mathcal{H}_2=\left|(\partial_{j} +iA_{j}) \varphi \right|^2
+\frac{1}{4}C^{a}_{j}C^{b}_{j}\left(\phi^a\phi^b -\phi^c\phi^c \delta^{ab} \right), 
\label{quadratic}
\end{align}
where we specialized the model to $N=3$ ($a=1,\ 2$) and defined the charged scalar  
and the $U(1)$ gauge field by
\begin{align}
\varphi=\frac{1}{2}(\phi^1+ i \phi^2),~~ A_j = \frac{1}{2} A^{ab}_j \varepsilon^{ab},
\label{varphi}
\end{align}
with $\varepsilon^{12}=-\varepsilon^{21}=1$.
A smooth $O(2)$ rotation of the local frame $\{\bm{e}^a\}$ naturally induces the $U(1)$ gauge transformation for $\varphi$. 
The second term obtained for the generic slow mode $\bm{n}_0$
also reduces to a nice geometric object once the BPS configuration for the slow modes $\bm{n}_0=\bm{n}_s$ 
satisfying
\begin{align}
\partial_{i}\bm{n}_s = \mp \varepsilon_{ij} \bm{n}_s\times \partial_{j}\bm{n}_s,
\end{align}
is adopted. This is equivalent to a Cauchy-Riemann (CR) look-alike duality relation
\begin{align}
C^{a}_i=\pm \varepsilon^{ab}\varepsilon_{ij} C^{b}_j,
\label{BPS0}
\end{align}
if the orientation convention $\bm{n}_s\times \bm{e}^b=-\varepsilon^{ab} \bm{e}^a$ is used.
 The plus and the minus signs in \eqref{BPS0}
 correspond to the self-dual and anti-self-dual solitons, respectively.
The duality relation yields a manifestly gauge invariant form
\begin{align}
\mathcal{H}_2=&\left|(\partial_{j} +iA_{j}) \varphi \right|^2
\pm  \frac{1}{2} \varepsilon_{ij}F_{ij}\ |\varphi|^2
\label{quadraticF},
\end{align}
with the field strength $F_{ij}$, 
which has also an interpretation as the topological charge density
\begin{align}
F_{ij}=&\partial_i A_j-\partial_j A_i=\varepsilon^{ab}C^{b}_{i}C^{a}_{j}\nonumber\\
=&-\bm{n}_s\cdot \partial_i\bm{n}_s\times \partial_j\bm{n}_s.
\label{nnn0}
\end{align}
Since this gives the Jacobian of the configuration map $x\in \mathbb{R}^2 \to  \bm{n}\in S^2=O(3)/O(2)$,
the integration over the whole plane yields
\begin{align}
\int {d}^2x~ \varepsilon_{ij}F_{ij}=-8\pi q,
\label{topological8piq}
\end{align}
where we used Vol$(S^2)=4\pi$ and the topological charge $q$ counts the wrapping number of the map
as $x$ sweeps over the base plane. 
 
\subsection{Effective magnetic field associated with solitons 
derived from the $\mathrm{CP}^1$ formulation}
\label{section:effective}

 We now consider the $\mathrm{CP}^1$ formulation, which is equivalent to
 the $O(3)$ formulation above, but is more convenient
 for dealing with the explicit form of the soliton configurations, while
we find the derivation of the quadratic part \eqref{quadraticF}
is most transparently performed in the $O(3)$ formulation
\footnote{To derive \eqref{quadraticF} from $\mathrm{CP}^1$, one should also expand $\tilde{A}_j$  
in \eqref{tildeA} in fluctuations, which is more complicated than it sounds.}.
The same energy in \eqref{H_O3} is now expressed as
\begin{align}
H=&\int {d}^2x~\left( |\partial_{j}\bm{z}|^2+|\bm{z}^{\dagger}\partial_{j}\bm{z}|^2 \right) \nonumber\\
=&\int {d}^2x~|(\partial_j +i \tilde{A}_j) \bm{z}|^2,
\label{H_CP1}
\end{align}
where the basic degree of freedom is the spin-$1/2$ (2-component) spinor $\bm{z}$ with $|\bm{z}|=1$,
which is related to the $N=3$ vector by the Hopf map $\bm{n}_\ell=\bm{z}^{\dagger}\sigma_\ell \bm{z}$ 
with $\sigma_\ell$ being the Pauli matrices for $\ell=1,2,3$,
and the gauge field in  $\mathrm{CP}^1$ is given by
\begin{align}
\tilde{A}_j=i\bm{z}^{*}\cdot \partial_{j}\bm{z}=\frac{i}{2}(\bm{z}^{*}\cdot \partial_{j}\bm{z}-\bm{z}\cdot\partial_{j}\bm{z}^{*}).
\label{tildeA}
\end{align}
It is worth recognizing a subtle difference between the gauge field here and the one \eqref{varphi}  
introduced naturally in the $O(3)$ formulation.  By using the Hopf map, one has
\begin{align}
A_j=2\tilde{A}_j +\partial_j\Lambda,
\label{A_tildeA}
\end{align} 
with some smooth function $\Lambda$.
They are thus related by factor two modulo gauge transformation. 

In terms of the covariant derivative $\tilde{\mathcal{D}}_j =\partial_j +i \tilde{A}_j$,
the finiteness of the energy \eqref{H_CP1} requires the asymptotic behavior $\tilde{\mathcal{D}}_j\bm{z}(x)=0$ as $|x|\to \infty$  and then $\bm{z}(x)\to \bm{z}_0 e^{i\phi}$ with some constant 2-component modulus $\bm{z}_0$  
and the phase variable $\phi$.
The single-valuedness of $\phi$ at the spatial infinity ($x=re^{i\theta}$ with $r\to \infty$) 
naturally defines the topological charge $q$
as the winding number of 
the map: $\theta\in S^1 \to \phi\in S^1$.  
This is given by $2\pi iq=\int_0^{2\pi} {d}\theta~  i\frac{{d}\phi}{{d}\theta}$,
and by noting \eqref{tildeA} and $\tilde{\mathcal{D}}_j\bm{z}=0$ as $r\to \infty$ , one obtains
\begin{align}
2\pi i  q=&-\oint dx_j~ i\tilde{A}_{j} 
=-i\int {d}^2x \varepsilon_{ij}\partial_i\tilde{A}_{j}\nonumber\\
=&\int {d}^2x\varepsilon_{ij}(\tilde{\mathcal{D}}_i\bm{z})^* (\tilde{\mathcal{D}}_j\bm{z}).
\label{topological4piq}
\end{align}
The last relation along with a simple observation
\begin{align}
0 \leqslant& \int {d}^2x~\frac{1}{2}|(\tilde{\mathcal{D}}_i\pm i\varepsilon_{ij}\tilde{\mathcal{D}}_j)\bm{z}|^2\nonumber \\
=&\int {d}^2x~
\left[(\tilde{\mathcal{D}}_j\bm{z})^* (\tilde{\mathcal{D}}_j\bm{z})\pm 
i\varepsilon_{ij}(\tilde{\mathcal{D}}_i\bm{z})^* (\tilde{\mathcal{D}}_j\bm{z})
\right],
\end{align}
leads to the Bogomol'nyi inequality   
\begin{align}
H\geqslant 2\pi |q|,
\label{Bogomolnyi}
\end{align}
which gives the lower energy bound for each topological sector.
The inequality is saturated with $\pm q\geqslant 0$ when one of the self-duality conditions  
$0=(\tilde{\mathcal{D}}_i \pm i\varepsilon_{ij} \tilde{\mathcal{D}}_j)\bm{z}$ is satisfied, where the plus and the minus signs
correspond to the self-dual and anti-self-dual solitons, respectively. 
Hereafter, we just refer them as the solitons ($q>0$) and anti-solitons ($q<0$).
The use of the stereographic coordinate for $\bm{n}$ denoted by $\bm{w}=\frac{\bm{z}}{z_1}$, further reduces these conditions to
the CR relation \cite{belavin1975metastable}
\begin{align}
0=\partial_i\bm{w} \pm i\varepsilon_{ij} \partial_j\bm{w},
\label{cauchyriemann}
\end{align}
and soliton configurations can be constructed 
\footnote{Note the distinction between the bold $\bm{z}$ (spinor) and the ordinary $z$ (position).
We choose $z_1$ to be non-vanishing component of  $\bm{z}$.}
from the holomorphic function in the complex coordinate $z=x_1+ix_2$.
With the standard basis $\bm{u}=(1,0)$ and $\bm{v}=(0,1)$, 
a charge $q>0$ soliton configuration may be constructed as
\begin{align}
\bm{w} =\bm{u}+\bm{v} W(z), \quad   W(z)=\prod_{j=1}^{q} \frac{(z- \hat{z}_j)}{\ell_j}, 
\label{solitonconfiguration}
\end{align}
where $\{\hat{z}_j , \ell_j\}\in \mathbb{C}^2$ are the center position and 
the size-parameter 
of each soliton, respectively. For more general form, see the remark at the end of this subsection.
One can use  \eqref{cauchyriemann} in \eqref{tildeA} to obtain the effective gauge field
\begin{align}
  \tilde{A}_j =& \frac{i}{2|\bm{w}|^2}(\bm{w}^{*}\cdot\partial_j \bm{w}-\bm{w}\cdot\partial_j \bm{w}^{*}) \nonumber\\
  =&\pm \frac{1}{2}\varepsilon_{jk}\partial_k \ln |\bm{w}|^2,
  \label{tildeA_lnw}
\end{align}
and evaluate the corresponding magnetic field 
that appears in the magnon Hamiltonian \eqref{quadraticF}:
\begin{align}
B_3 \equiv\frac{1}{2}\varepsilon_{ij}F_{ij}= 2 \varepsilon_{ij}\partial_{i} \tilde{A}_j,
\label{Bz_def}
\end{align}
where the factor $2$ in \eqref{A_tildeA} is taken into account. 
As the magnetic field in 2+1d has only one component, 
the formal subscript 3 is temporarily introduced just to remind us that 
its sign depends on the sign of $q$; our convention, $B_3<0$ $(B_3>0)$ for the solitons $q>0$
(anti-solitons $q<0$), is the same as that used in Ref. \onlinecite{rajaraman1982solitons} 
(which is opposite to Ref. \onlinecite{d19781n}).
In order to construct anti-soliton configurations, 
one may use functions 
of an anti-holomorphic variable
$\bar{z}=x_1-ix_2$ instead of $z$ in \eqref{solitonconfiguration}.

It is convenient to introduce the derivatives with respect to the complex coordinates $z$ and $\bar{z}$
defined respectively by $\partial=\partial_{z}=\frac{1}{2}(\partial_{x_1} - i\partial_{x_2})$
and $\bar{\partial}=\partial_{\bar{z}}=\frac{1}{2}(\partial_{x_1} + i\partial_{x_2})$,
with which the CR relation \eqref{cauchyriemann} simplifies
to $\bar{\partial}{W}=0$ for $q>0$ (or ${\partial}{W}=0$ for $q<0$). 
This also means a configuration derived from \eqref{solitonconfiguration} with a function $W(z, \bar{z})$ which depends on both $z$ and $\bar{z}$ does not saturate the Bogomol'nyi inequality.
In particular, it is important to recognize that a soliton-anti-soliton pair in the $q=0$-sector
interacts even at classical level.

By contrast, we are interested in purely quantum interaction; 
this may be extracted from the configurations with 
$|q|>0$ where the Bogomol'nyi inequality is saturated so that either one of these CR relations holds. 
It is our aim to study the possible quantum interaction in the case where 
the classical interaction vanishes exactly.
Combining either one of the CR relations, \eqref{tildeA_lnw} and \eqref{Bz_def}
then yields 
a useful expression
\begin{align}
B_3 = - \varsigma \frac{4 \partial W \bar{\partial} \overline{W} }{(1+W\overline{W})^2},
\label{BWW}
\end{align}
with $\varsigma=\mathrm{sign}(q)$.
This particular form with $\mathrm{sign}(q)$ is also understood from a more general expression 
of the  topological charge density \cite{gross1978meron}
for $W=W(z, \bar{z})$:
\begin{align}
B_3 =  4\frac{ |\bar{\partial}W|^2 -|\partial W|^2 }{(1+W\overline{W})^2},
\end{align}
once one of the CR relations is used.
For the purpose we have just stated, however, it is enough to use \eqref{BWW} with 
$W=W(z)$ for $q>0$ (or $W=W(\bar{z})$ for $q<0$).

\begin{figure}
\includegraphics[width=7.5cm]{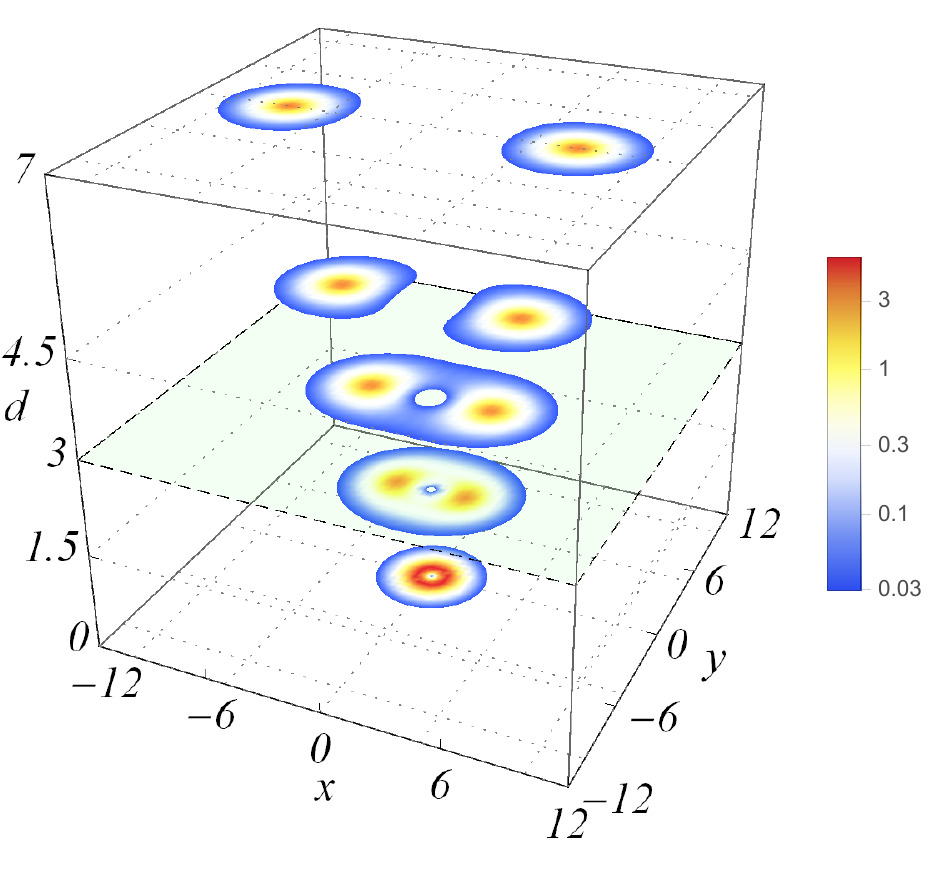} 
\caption{\label{fig:distance}
The magnetic field strengths $|B_3(x)|$ in \eqref{B_2soliton} 
for 2-solitons of the same size $\lambda=1$ 
with various distances $d=0, 1.5, 3, 4.5, 7$. The region for each $d$ with $|B_3(x)|<0.03$ is excluded.
They are identical to the topological charge densities of the optimal solitons 
that saturate the Bogomol'nyi bound for $q=2$. 
The density for $d\gg \lambda$ is asymptotically a superposition 
of two independent solitons in \eqref{B_1soliton} each shifted by $\pm d$,
while it is deformed as $d$ decreases and becomes eventually the ring configuration
in \eqref{B_nsoliton} with $n=2$ at $d=0$.
} 
\end{figure}

\begin{figure*}
\includegraphics[width=14.5cm]{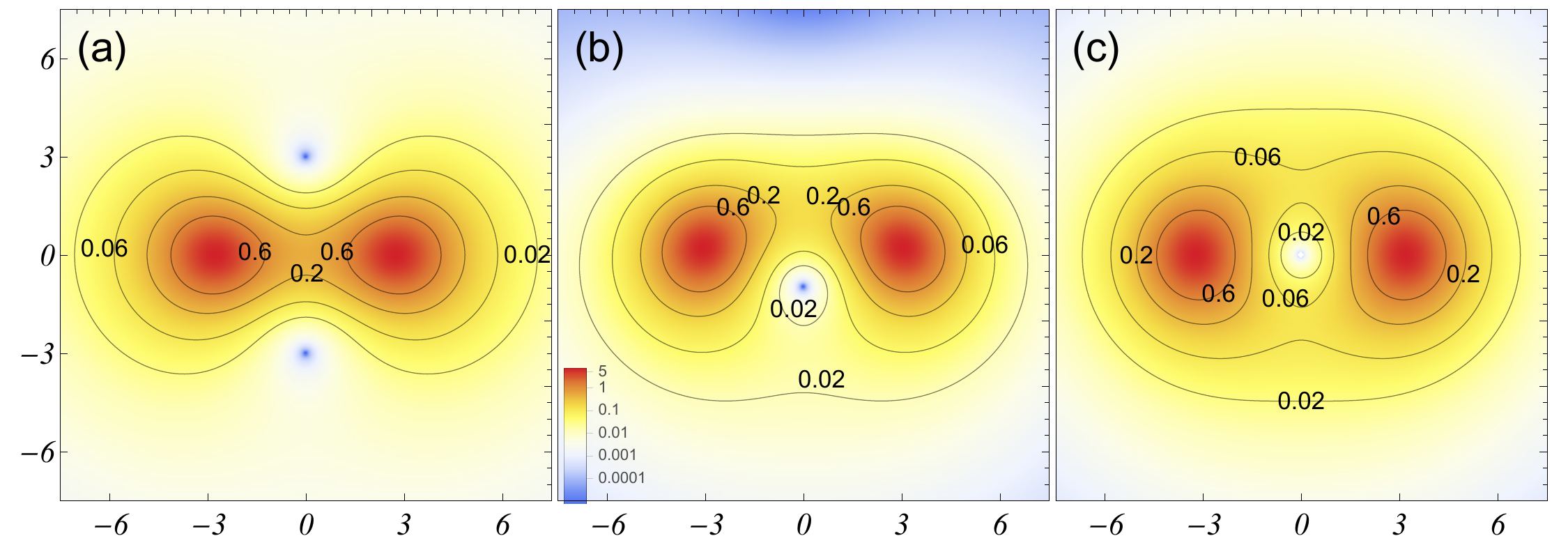} 
\caption{\label{fig:2solitons}
The magnetic field strength $|B_3(x)|$ for 2-solitons of the same size $\lambda=1$ 
each located at $d=3$ apart from the origin with various  phase angles
(a) $\theta=0$,
(b) $\theta=3\pi/5$,
(c) $\theta=\pi$, given by \eqref{BWW} and \eqref{Wz=},
which also equals to the topological charge density 
that saturates the Bogomol'nyi bound.
Classically, there are no forces between the solitons as they are optimized.
} 
\end{figure*}

More concretely, from \eqref{solitonconfiguration} and \eqref{BWW}, one has, 
\begin{align}
B_3  (x)= - \varsigma \frac{4\lambda^2}{\left(  \lambda^2+x_1^2+x_2^2\right)^2},
\label{B_1soliton}
\end{align}
with a linear map $W=z/\lambda$ representing a single soliton of size $\ell_1=\lambda$ at $\hat{z}_1=0+0i$. 
With a quadratic polynomial $W=(z^2-d^2)/{\ell^2}$, one has 
\begin{align}
B_3  (x)= - \varsigma \frac{16\ell^4 (x_1^2+x_2^2)}{\left(\ell^4+\left[(x_1 -d)^2+x_2^2][(x_1 +d)^2+x_2^2\right]\right)^2},
\label{B_2soliton}
\end{align}
with $\ell=\ell_1=\ell_2$ representing two solitons, one at $\hat{z}_1=d+0i$ and the other at $\hat{z}_2=-d+0i$ each with the same size $\lambda$ defined as the positive root of $\lambda(\lambda+2d)=\ell^2$.
Note that the shape of each soliton in \eqref{B_2soliton} is not a disk, but is deformed  as it is optimized classically
to saturate the bound \eqref{Bogomolnyi}.
This is plotted in FIG. \ref{fig:distance} (see also FIG. \ref{fig:2solitons}, where one may observe the shape of the Cassini ovals).
It is also of interest \cite{walliser2000casimir, ivanov2007quantum} to consider 
a $q=n$ soliton for the concentric configuration $\hat{z}_1=\cdots=\hat{z}_n=0+0i$ 
all with the same size; this is realized by
$W=\left({z}/{\lambda}\right)^n$, which is a zero of order $n$. 
From this, one obtains \cite{rajaraman1982solitons} 
\begin{align}
B_3  (x)= - \varsigma \frac{ 4n^2 \lambda^{2n} \left(x_1^2+x_2^2\right)^{n-1}}{\left(  \lambda^{2n}+\left(x_1^2+x_2^2\right)^n \right)^2},
\label{B_nsoliton}
\end{align}
which reduces to \eqref{B_1soliton} at $n=1$. 
For $n>1$, this corresponds to a ring configuration\cite{manton2004topological} taking the maximum at the radius $r=\lambda  \left(\frac{n-1}{n+1}\right)^{\frac{1}{2 n}}$.
Also \eqref{B_nsoliton} at $n=2$ is naturally a coincidence limit $d=0$ of the 2-soliton configuration \eqref{B_2soliton}.

As a brief remark, the magnetic field may depend on the relative orientations of the solitons.
To see this in 2-soliton case, note first that the topological degree of the rational map $W(z)=P(z)/Q(z)$ is given by $q=\max[\deg(P), \deg(Q)]$ if $P(z)$ and $Q(z)$ are two coprime polynomials.
For $q=2$ sector, 
the general configuration is thus given by $W=(\gamma_1 z +\gamma_2)/(z^2+\delta z +\varepsilon)$
with 4 complex parameters $\{\gamma_1, \gamma_2, \delta, \varepsilon\}$.
Since the overall phase of $W$ does not change the magnetic field \eqref{BWW}, 
the effective dimension of the real parameter space is reduced by one.
By fixing the center of mass of the two solitons and aligning them along the real axis, one has
$\delta=0$ and $\varepsilon\in \mathbb{R}_{<0}$.
We consider the symmetric case where the two solitons have the same size; 
this further reduces the dimension by one.
Such a 2-soliton configuration is then specified by 3 real parameters. A convenient representation may be
\begin{align}
W(z)=\left(\frac{\gamma_0}{z-d}+ e^{i\theta} \frac{\gamma_0}{z+d} \right), 
\label{Wz=}
\end{align}
with $\{\gamma_0, \theta, d\}\in \mathbb{R}^3$, where 
$\gamma_0=\ell^2/(2d)$ with $\ell^2=\lambda(\lambda+2d)$ is the size parameter,
and $2d$ is the distance between the solitons.
The magnetic field for various phase angle $\theta$ is plotted in FIG. \ref{fig:2solitons}.
The quantum interaction at short-range ($d\sim \lambda$) may be modified by the variation of 
the angle $\theta$; this may be related to the stability angle $\theta=\theta_*$ 
with which the 2-solitons becomes the most stable.
On the other hand, the long-range interaction at the leading orders in ${\lambda}/{d}$ remains the same in the analysis in Section \ref{section:quantum}.
Therefore, in this paper, we concentrate on the case with $\theta=\pi$, where the analytic expression in the polar-coordinates simplifies by symmetry.
This corresponds to the oppositely aligned configuration given by
the magnetic field \eqref{B_2soliton} plotted in FIG. \ref{fig:2solitons}-(c).

\subsection{The charge that couples magnon to soliton}
In the 2d quadratic Hamiltonian \eqref{quadraticF},
there is a symmetry between the soliton sectors ($q>0$) and the anti-soliton sectors ($q<0$).
This follows from the fact that 
the overall $\pm$ sign of the last term in \eqref{quadraticF} and 
the sign of the field strength $B_3 =F_{12}$ have the same origin. 
Thus, the Hamiltonian \eqref{quadraticF} take the following form
\begin{align}
\mathcal{H}_2=&\left|(\partial_{j} +ieA_{j}) \varphi \right|^2
-  e |B_3 | |\varphi|^2
\label{quadraticBz},
\end{align}
which depends on the absolute value $|B_3|$.

The charge $e>0$ just introduced in \eqref{quadraticBz}
can be thought of as an effective charge that couples the magnon and the magnetic field induced by the soliton configuration.
In the classical analysis based on the $O(3)$ model, one has charge $e=1$ as we do in \eqref{quadraticF}.   

To avoid possible confusions, it is important to note that one would have charge two ($\tilde{e}=2$) if one had used the $\mathrm{CP}^1$ gauge field $\tilde{A}_j$ in \eqref{quadraticBz} instead of the $O(3)$ gauge field $A_j$.
The ratio $\tilde{e}/e=2$  would be understood 
from \eqref{A_tildeA} due to  
the vector-spinor relation given by the Hopf map $\bm{n}_i=\bm{z}^{\dagger}\sigma_i \bm{z}$,
or more globally from the ratio 
between the natural units of the total flux, Vol$(S^2)$ in \eqref{topological8piq} and 
Vol$(S^1)$ in \eqref{topological4piq}.

It is important to realize that the straight-forward classical analysis \footnote{
The situation is the same at the quantum level
as the original $O(3)$ symmetry is preserved under renormalization.
Also, the charge renormalization in QED does not affect our calculation in Section \ref{section:quantum} where 
the effective potential is a function of the invariant combination $eB$. } 
of the energy \eqref{H_O3} leads to only single charge $e$.  
It is, however, instructive to further generalize \eqref{quadraticBz}
by introducing a redundant parameter $\kappa$.
Hereafter we also write $B=B_3>0$ instead of $|B_3|$ for simplicity;  one may study the case with $q<0$ without loss of generality.
This leads to the magnon-soliton coupling
\begin{align}
\mathcal{H}_2=&\left|(\partial_{j} +ieA_{j}) \varphi \right|^2
-  \kappa e B |\varphi|^2.
\label{k_quadraticB}
\end{align}
In Section \ref{section:relation}, we discuss how to deal with this coupling 
in view of the resemblance to that in the non-relativistic limit of the QED.

We find the right choice $\kappa=1$ significantly simplifies the quantum analysis.
In order to backup the main computation in \ref{section:derivative},
this aspect is emphasized in Section \ref{section:relation} 
and more concretely in Section \ref{section:spectral}.
A practical reader could directly go to Section \ref{section:quantumfluctuation} just by keeping in mind that 
the geometric derivation has lead us to $e=1$ and $\kappa=1$ as in \eqref{quadraticF}.

As a technical remark, the form of the magnon-soliton coupling in \eqref{k_quadraticB} with $e=\kappa=1$
is essentially known in the first quantized picture, namely, in the context of 
single magnon scattering off the soliton \cite{ivanov2007quantum},
where our second term $-\kappa e B |\varphi|^2$ exactly corresponds to
the potential $V$ for a magnon in Eq. (3a) of Ref. \onlinecite{ivanov2007quantum}.
In a special circumstances, it is known that the magnon scattering may be treated 
in an AB-scattering approximation where this potential term is omitted\cite{iwasaki2014theory};
this corresponds to $\kappa=0$.
It is, however, worth emphasizing that this term could play an important role in general 
as shown in Section \ref{section:spectral}.

\subsection{Relation to the scalar/spinor QED and the non-relativistic limit: the Pauli Hamiltonian}
\label{section:relation}

The 2d Hamiltonian \eqref{k_quadraticB} is closely related 
to the fermionic functional determinant since the squared Dirac operator in 2d is given by
\begin{align}
\slashed{D}^2= 
\left(\partial_{j} +ieA_{j}\right)^2 + \frac{e}{2}\sigma_3 \varepsilon_{ij}F_{ij} 
\label{diracsquared},
\end{align}
where the spinor components are distinguished by $\sigma_3=\mathrm{diag}(1,-1)$.
This analogy leads us to take the strategy in Section \ref{section:worldline} as follows.

The effective action which arises from $\slashed{D}$ for various $d$,
namely, that of the spinor QED (including the standard QED$_{3+1}$) in an external field
can be computed by the worldline formalism\cite{strassler1992field, schmidt1993calculation, schubert2001perturbative},
where the first quantized picture of the spinor particle 
is realized by introducing auxiliary Grassmann coordinates.
In contrast, we note that the Hamiltonian \eqref{k_quadraticB} has no such spinor components; 
a projection to the 
state by the rule $\sigma_3\to \mathrm{sign}(F_{12})$ would be necessary to relate them. 
For this reason, we adopt the worldline formalism for the spinless particle  (scalar QED) as a basis of the computation.

In such contexts, the effective action is closely linked to a certain non-relativistic one-particle Hamiltonian. 
For instance, the QED$_{3+1}$ effective action
in a uniform magnetic field is essentially determined \cite{brezin1970pair} 
 from the eigenvalues of the following type: 
$E=\mathcal{E}+E_{n,\pm}$ with 
\begin{align}
\mathcal{E}=\frac{p_\perp^2}{2m}, \quad E_{n,\pm}=\frac{eB}{m}\left(n+\frac12 \pm \frac{\kappa}{2} \right),
\label{landaulevel}
\end{align}
where $p_\perp^2 =-p_0^2 + p_z^2$ is the squared momentum transverse to the cyclotron plane.
One would recognize a 3+1d version of the Landau levels in \eqref{landaulevel},
which generalize the levels in the Pauli Hamiltonian for a non-relativistic spin-$1/2$  particle: 
\begin{align}
\hat{H}=\frac{(\bm{p}-e\bm{A})^2}{2m}    -   \mu\, \bm{\sigma}\cdot \bm{B},  \quad \mu=\kappa \frac{e}{2m}, 
\label{pauliNonrela}
\end{align}
in a uniform magnetic field, with which one has just $p_\perp^2=p_z^2$.
One recognizes that the redundant parameter $\kappa$
is the half of the gyromagnetic ratio in the magnetic moment 
$\mu$.
Were it not for the radiative corrections,  
one has exactly $\kappa=1$, leading to the $\mathcal{N}=2$ supersymmetry \cite{cooper1995supersymmetry}
manifested in the twice degenerate levels\cite{landau1958quantum, weinberg2015lectures} 
above the isolated lowest Landau level $E_{0, -}$. 
The level spacing at $\kappa=1$ indeed simplifies our results as corroborated in Section \ref{section:spectral}.

The spinor QED effective action derived from \eqref{diracsquared} corresponds to $\kappa=1$,
while the scalar QED effective action corresponds to $\kappa=0$.
According to \eqref{quadraticF} and \eqref{quadraticBz}, the soliton-magnon system corresponds to 
the {minus} branch $E_{n, -}$ 
of the levels \eqref{landaulevel} at $\kappa=1$.
In a sense, the soliton-magnon system may be seen as a deformation of the scalar QED 
exactly up to the point where it resembles a half of the spinor QED.
This connection is corroborated by evaluating the spectral zeta function for general $\kappa$
in Section \ref{section:spectral},
where the effective action for
the constant magnetic field case is derived as an
independent check of the derivative expansion obtained Section \ref{section:derivative}.

\section{Quantum Interaction from the Worldline effective action} \label{section:worldline}

\subsection{Quantum fluctuation around the static solitons}
\label{section:quantumfluctuation}
We consider 2d isotropic spin-$S$ antiferromagnets described by the quantum Hamiltonian 
$H=J\sum_{\langle ij \rangle}{S_i \cdot S_j}$, where $J>0$ is the exchange coupling and 
the sum is over the nearest-neighbor spins.
At low temperatures, the long wavelength physics is described by
the N\'{e}el sublattice magnetization denoted by a unit vector $\bm{n}$ 
obeying the action, 
closely related to \eqref{H_O3}, of the 2+1d O(3) non-linear sigma model 
augmented by the Berry-phase term\cite{haldane1988monopole,sachdev2007quantum}
\begin{align}
\mathcal{S} =& \mathcal{S}_B + \int_{0}^{\mathcal{T}}\!\!\! dt \int d^2 x~ \frac{\rho_s}{2} \left[ (\partial_0 \bm{n})^2 - (\partial_j \bm{n})^2 \right], 
\label{S=SB}
\end{align}
where 
$\rho_s$ is the spin stiffness and $\partial_0=c_s^{-1}\partial_t$ is defined along with the spin-wave velocity $c_s$. 
Their bare values are given by $\rho_s=JS^2/\hbar$ and $c_s=2\sqrt{2}JSa/\hbar$ with the lattice constant $a$.
Hereafter, for the time being, we set $\rho_s=1/2$ and $c_s=\hbar=1$.
 The term $\mathcal{S}_B$ unspecified here is omitted 
below as it vanishes for the simplest cases in antiferromagnets.
For instance, on a square lattice,  the contributions of $\mathcal{S}_B$ from two sublattices cancels out with each other,
because of the alternating nature of the N\'{e}el state,  
for smooth configuration of $\bm{n}$, which is relevant for our problem.

Note that even if we assume a smooth spin configuration, it is not obvious that the Berry phase can be omitted for the fluctuation as it may be ``very rough" on the lattice.
This is not an issue if we can restrict our path-integral to the local fluctuation, as we do, by excluding the large fluctuation corresponding to the histories that become singular in the continuum limit.  Such a restriction may be justified in the ordered phase since the smooth histories of the configuration dominate over the singular histories. 

In a broader perspective, a singular histories $\bm{n}(t,x)$ arising from the fluctuation in 2+1d space-time 
 known as a hedgehog  may describe a tunneling event between different topological sectors and may contribute non-trivially in the path-integral via the Berry phase \cite{haldane1988monopole,read1990spin}. Here we are, however,  interested in possible quantum interactions in the solitons of size much larger than the lattice spacing  $a$ with a fixed topological charge $q$ by focusing on the ordered phase.
The path-integral is then dominated by smooth histories of $\bm{n}(t,x)$, for which  the Berry-phase term can be ignored \cite{haldane1988monopole,fradkin1988topological,wen1988spin,dombre1988absence,sachdev2007quantum}
 although it could be important in the disordered phase \cite{read1990spin}\footnote{The Berry phase could play a role for the virtual processes of the topological sector tunnelings such as $q \to (q\pm 1) \to q$, which are suppressed in the ordered phase.}, which is beyond the scope of this paper.

 We further focus on the quantum fluctuations around the static soliton configurations satisfying
 $\partial_0 \bm{n}_s=0$.
 Using \eqref{quadratic} and \eqref{k_quadraticB},
the quadratic Lagrangian would then be given by 
\begin{align}
\mathcal{L} =& \partial_0\varphi^*\partial_0\varphi - \mathcal{H}_2 -m^2|\varphi|^2 \nonumber \\
=&\left|\mathcal{D}_{\mu}\varphi\right|^2 + (\kappa e B -m^2)\left|\varphi\right|^2,
\label{quadratic_L}
\end{align}
with the covariant derivative
$\mathcal{D}_{\mu}=\partial_{\mu}+ieA_{\mu}$ and the vanishing time component $A_0=0$.
The effective magnetic field $B=|F_{12}|$ along with the gauge field $A_i$ $(i=1,2)$ is induced by 
the solitons saturating the Bogomol'nyi inequality, 
and correspond to \eqref{B_1soliton} or \eqref{B_2soliton} for instance.
Although \eqref{quadratic_L} has three parameters $\{e, \kappa, m\}$, 
one should eventually take $e=\kappa=1$ as in \eqref{quadraticF}
and take the limit $m\to 0$ 
so that 
the action reduces to \eqref{S=SB} 
describing the massless spin wave above the ordered state
\cite{sachdev2007quantum} 
once the effective magnetic field due to the background solitons is turned off;
as is well-known, 
this massless property could be traced back to the existence of the degenerate 
spin configurations in the trivial sector $q=0$.

Adding up the fluctuations around the solitons for the non-trivial topological sector ($|q|>0$), in principle, needs some reflections on the path-integral measure (and the resulting functional determinant)
since there are zero-modes associated with the flat-directions, as we now see, 
which do not affect the quantum physics of the solitons though. 
These flat directions form the moduli (parameter) space of solitons due to the exact degeneracy of the classical energies of the soliton configurations saturating the Bogomol'nyi inequality.
In our case, the moduli space is $4q$-dimensional in real coordinates \cite{manton2004topological}. 
For one soliton ($|q|=1$), we have two freedoms for the positions and the other two for the sizes and the phases, which correspond to the zero-energy magnon fluctuations\cite{ivanov2007quantum} associated with the translational, dilatational (if the inequality is saturated), and rotational symmetry of the soliton, or equivalently, the magnetic skyrmion. 

To deal with such flat directions, which introduce apparent divergences in the Gaussian integral, one may use the collective coordinates method\cite{gervais1976collective, thooft1976computation}, developed as the Fadeev-Popov procedure for fixing the soliton configurations to one particular point in the moduli space.
The integration over the zero modes is then transformed into an integration over the collective coordinates; in the partition function $Z$, this would yield the Jacobian factor proportional to the norm $J_\ell\propto \left(\int d^2x\ |\bm{n}_{s; \ell}'|^2\right)^{1/2} $ for the unnormalized zero eigenmode $\bm{n}_{s; \ell}'$ obtained as the derivative of the soliton solution in the $\ell$-th flat direction. For translational modes, it is the square root of the classical energy, which is independent of the choice of the moduli (of the center of the solitons, in particular). 
The net factor thus only gives rise to an unimportant offset to the effective potential, which has no effect on the solitons in this setting.
Thus, we hereafter fix the soliton configurations and omit the zero modes;
in practice, one should replace the functional determinant $\det \mathcal{X}$ of the quadratic operator $\mathcal{X}$ by $\det' \mathcal{X}$, a regularized infinite product of all the eigenvalues but the zeros. In particular, the Gaussian integral just evaluated below is given by $\mathrm{tr} \ln \mathcal{X}\equiv \ln\det' \mathcal{X}$ although the zero eigenvalues corresponding to the original flat directions do not arise in the following analysis as our derivative expansion is around the constant field $B$.    

\subsection{Worldline formulation}
Let us consider the correction to the vacuum energy due to the magnon-soliton coupling \eqref{quadratic_L}
in the non-trivial sectors.
Since the partition function behaves as $Z=\langle 0| e^{-i\mathcal{H}\mathcal{T}}|0 \rangle\to  e^{-iE_0 \mathcal{T}}$ 
for the late times $\mathcal{T}\to \infty$, the effective potential can be obtained by $V=-\frac{1}{\mathcal{T}}S_{\mathrm{eff}}$ with the definition $Z=e^{iS_{\mathrm{eff}}}$.
The partition function of the magnon-soliton system is given by
\begin{align}
Z=&\int \mathcal{D}\varphi~ e^{i \int_0^\mathcal{T} dt \int d^2x~ \varphi^*\left(-\mathcal{D}_{\mu}\mathcal{D}^{\mu} + \kappa e B -m^2\right)\varphi},
\label{Z=}
\end{align}
which yields after the Gaussian integral
\begin{align}
-\mathcal{T}V 
=-i\ln Z= i \mathrm{tr} \ln \left(-\mathcal{D}_{\mu}\mathcal{D}^{\mu} + \kappa e B  -m^2\right).
\label{TV=minus_i}
\end{align}
In order to obtain a tractable expansion,
the key step is writing this in the worldline formalism \cite{schubert2001perturbative} 
\begin{align}
 \mathcal{T} V= i \int_{0}^{\infty} \frac{dT}{T} ~ e^{ im^2 T}\mathrm{tr}\ e^{-i(H+\kappa eB)T}
 \label{TV=}
\end{align}
where the proper-time length $T$ of the magnon loop  is introduced using 
the identity
\begin{align}
\ln \mathcal{X}/\mathcal{X}_0= \int_{0}^{\infty} \frac{dT}{T}\left(e^{-i (\mathcal{X}_0-i\varepsilon) T}-e^{-i (\mathcal{X}-i\varepsilon) T} \right)
\label{ln_identity}
\end{align}
 with $\mathcal{X}=H+ \kappa eB - m^2 $ and $H=-\mathcal{D}_{\mu}\mathcal{D}^{\mu}
 =p_0^2-(p+eA)^2$. 
The free theory contribution from $\mathcal{X}_0=-\partial_{\mu}\partial^{\mu} - m^2 $ is omitted below as it is not important.
  Then the trace for one-particle, non-relativistic Hamiltonian $H$ can be evaluated as
\begin{align}
\mathrm{tr}&\ e^{-i(H +\kappa eB)T}  
= \int_{x(0)=x(T)} \hspace{-10mm} \mathcal{D}x \mathcal{D}p~
 e^{i\int_0^T ds\  \left( p\dot{x} - H -\kappa eB\right)} \nonumber \\
&=\mathcal{N}\mathcal{T}\!\!
\int d^2 \bar{x}\,  e^{-i\kappa eB(\bar{x})T}
\!\! \int_{\bar{x}=x(0)=x(T)} \hspace{-15mm} \mathcal{D}x~
 e^{i\int_0^T ds\, L\left(x(s),\, \dot{x}(s) \right)} 
\label{U=int}
\end{align}
 where $\mathcal{N}$ is a normalization factor \cite{strassler1992field} 
accounting the Gaussian momentum path-integral $\int \mathcal{D}p$, and
 $\bar{x}$ is a reference point passed by the magnon loop. 
 The Lagrangian is given by 
 \begin{align}
 L\left(x,\, \dot{x} \right)= \frac{\dot{x}^2}{4} +e A_\mu(x) \dot{x}^\mu - 
 \kappa e\left(B (x)- B (\bar{x}) \right),
 \label{L=}
 \end{align}
where 
the subtraction of $B(x)$ by a constant value $B(\bar{x})$ is due to the decomposition in \eqref{U=int}, 
and does not affect the equation of motion.
This reduces to the Lagrangian for the scalar QED particle if one turns off the last term by setting $\kappa=0$.
The 
effective action for $\kappa = 0$ is essentially given in the worldline form \eqref{TV=} with \eqref{U=int}
in Ref. \onlinecite{feynman1950mathematical},
and is also computed exactly by the onshell-action solved by the equation of motion
for a constant field case \cite{nambu1950use, dunne2012heisenberg}. 

The main problem is then an evaluation of 
the return amplitude in \eqref{U=int} 
\begin{align}
\langle \bar{x}| U  |\bar{x}\rangle&\equiv
\int_{\bar{x} = x(0)=x(T)} \hspace{-15mm} \mathcal{D}x~
e^{i\int_0^T ds\  L\left(x(s),\, \dot{x}(s)
 \right)},
\label{return}
\end{align}
at a given proper-time loop length $T$.  
Here the evolution operator $U$  
 corresponding to \eqref{L=} is introduced for the notational convenience.

In the presence of an inhomogeneous field and a non-zero coupling $\kappa$, 
the full effective action may be expanded in powers of the derivatives of $B$.
To develop such an expansion, the most natural thing is to expand 
$A_\mu(x)$ and $B(x)$ in \eqref{L=}
in terms of the displacement from a reference point $\bar{x}$. 
This can be done by using the standard expansion \cite{pascual1984qcd}
\begin{align}
A_\mu(\bar{x} + x)  =& \sum_{n=0}^{\infty} \frac{x_\lambda x_{\nu_1}\cdots x_{\nu_n}}{n! (n+2)}
\partial_{\nu_1}\cdots \partial_{\nu_n}
F_{\lambda \mu} (\bar{x}),
\label{FS_gauge_A}
\end{align}
in the Fock-Schwinger gauge along with the expansion
\begin{align}
B(\bar{x} +x) - B (\bar{x}) =& \sum_{n=1}^{\infty} \frac{x_{k_1}\cdots x_{k_n}}{n!}
\partial_{k_1}\cdots \partial_{k_n}
B (\bar{x}). 
\label{FS_gauge}
\end{align}
Then the return amplitude \eqref{return} can be evaluated with the straightforward decomposition
$L = L_{0}+L_{1}$ with
\begin{align}
L_{0} =& \frac{\dot{x}_0^2}{4}-\frac{\dot{x}_j^2}{4} + \frac{e}{2}x_i F_{ij} \dot{x}_j,
\label{L0}\\
L_{1} =&
e x_i\dot{x}_j\sum_{n=1}^{\infty} \frac{ x_{k_1}\cdots x_{k_n}}{n! (n+2)}
\partial_{k_1}\cdots \partial_{k_n}
F_{ij} (\bar{x})  \nonumber \\
&-\kappa e \sum_{n=1}^{\infty} \frac{ x_{k_1}\cdots x_{k_n}}{n!}
\partial_{k_1}\cdots \partial_{k_n}
B (\bar{x}),
\label{L1} 
\end{align}
into the non-derivative terms $L_0$
and the derivative terms $L_1$. 
Here only non-zero components $F_{ij}=B\varepsilon_{ij}$ 
of the field strength are magnetic field;  
one has the zero electric field $F_{0j}=0$  for the static soliton configurations. 
 
According to \eqref{return}, one may consider the magnon as a usual charged particle 
in scalar QED as prescribed in \eqref{L0},
except that it feels the inhomogeneity of the soliton magnetic field as in \eqref{L1}. 
In order to obtain the lowest order terms which contain at most two derivatives,
it is sufficient to further decompose the derivatives in \eqref{L1} into two parts, 
and to truncate each of them as 
\begin{align}
\label{L1=} L_{1}=&L_{10}+L_{11}, \\
\label{L10} L_{10}=&\frac{e}{3}\varepsilon_{ij} x_i \dot{x}_j x_k \partial_k B + \frac{e}{8}\varepsilon_{ij} x_i \dot{x}_j x_k x_\ell \partial_k \partial_\ell B,    \\
\label{L11} L_{11}=& -\kappa e\ x_j \partial_j B - \frac{\kappa e}{2}  x_j x_\ell \partial_j \partial_\ell B,
\end{align}
where $L_{10}$ contains an inhomogeneity correction to the magnon AB-scattering due to 
the long-range soliton gauge field $A_\mu(x)$,
and $L_{11}$ takes account the variation of the magnetic field $B(x)$, 
which acts like an on-site potential obstruction to the magnon.
For the interplay between $eA_\mu(x)$ and $-\kappa eB(x)$,
see the last paragraph of Section \ref{section:spectral}.
In our worldline view, it is essential to include the quantum phase from the both terms.

\subsection{The derivative expansion for the soliton Casimir energy}
\label{section:derivative}
Let us denote the partial evolution operator for the homogeneous (constant field) problem by 
$U_c=e^{-iH_c T}$ for the (pseudo-)Hamiltonian $H_c=p_0^2-(p+eA)^2$ with
$A_\mu(x)=\frac{1}{2}(x_{\lambda}-\bar{x}_\lambda)F_{\lambda\mu}(\bar{x})$, which is
the $n=0$ term in \eqref{FS_gauge_A}.
This corresponds to the Lagrangian $L_0$ in \eqref{L0} for the constant field. 
Then the return amplitude in \eqref{return} may be written as
\begin{align}
\langle \bar{x}| U&  |\bar{x}\rangle=\langle \bar{x}| U_c  |\bar{x}\rangle \cdot \Xi, \label{U=UcXi}\\
\Xi =&\langle e^{i L_{1}} \rangle= 1 + i\int_0^T ds \langle L_{1}(s) \rangle  \nonumber\\
& -\frac{1}{2}\int_0^T\int_0^T ds_1 ds_2 \langle L_{1}(s_1)L_{1}(s_2) \rangle+ \cdots,
\label{Xi}
\end{align}
where the expectation values are with respect to the motion \cite{nambu1950use,gusynin1999derivative} 
along the proper-time loop determined from $L_0$ in \eqref{L0}.
The first term in the derivative expansion can be obtained from \eqref{U=int}
using the known result\cite{dittrich2000probing} (the normalization is fixed by considering the $T\to 0^-$ limit \cite{schwinger1951gauge})
\begin{align}
\mathcal{N} \langle x| U_c |x \rangle = \frac{e^{-\pi i/4}}{(4\pi T)^{3/2}}\frac{ieBT}{\text{sh}\ ieBT}
\end{align}
by setting $\Xi=1$ in \eqref{Xi}. This yields 
\begin{align}
V_c=&i\int_{0}^{\infty} \frac{dT}{T}\int d^2 \bar{x} e^{-i (\kappa  e B -m^2) T} \mathcal{N} \langle \bar{x}| U_c  |\bar{x}\rangle \nonumber\\
=& - \alpha \int d^2 x \left(\frac{eB}{4\pi}\right)^{3/2}\!\!\!,\ \ \alpha = \int_{0}^{\infty}\!\!\!\!\! dw~ w^{-\frac{3}{2}}  \frac{e^{- \nu w }}{\text{sh}\ w}, 
\label{V_c=}
\end{align}
where the combination $\kappa e B-m^2$ is to be associated with an infinitesimal imaginary part 
$-i\varepsilon$ in \eqref{ln_identity}, 
and the integration path is rotated accordingly so that effectively a new variable $w=ieBT$ is used.
We use the following value for the exponent 
\begin{align}
\nu=\kappa  - \frac{m^2}{eB} \to 1,
\label{nu=}
\end{align}
in the limit of physical interest, where the magnon mass $m$ vanishes.
The exact value of $\alpha$ is given shortly in \eqref{alpha=zeta} along with the next coefficient $\beta$. 

In order to compute the effect of the inhomogeneity as in \eqref{L10}-\eqref{Xi},
one needs to evaluate the expectation values of the fields $x_i(s)$ and $\dot{x}_i(s)$ $(i=1,2)$ originating from the magnon spatial coordinates; this is a 0+1d field theory (quantum mechanics) living on the proper-time loop, 
where the coordinates are dynamical variables. 
The subsequent analysis is based on the propagator \cite{gusynin1999derivative}    
\begin{align}
\langle x_i(s_1) x_j(s_2)\rangle=i g_{ij}(s_1, s_2),
\label{xixj}
\end{align}
(and its proper-time derivatives)
determined by the on-shell action 
solving the equation of motion from \eqref{L0} along the loop for the Dirichlet boundary conditions, yielding
\begin{align}
g_{ij}(s_1, s_2) = 
e^{-eFs_{-}}\frac{\text{ch}\: {eF(T-s_{+})} -\text{ch}\: {eF(T-|s_{-}|)}        }{eF\ \text{sh}\ eFT}  
\label{gij}
\end{align}
with $F=F_{ij}=B\varepsilon_{ij}$ and $s_{\pm}=s_1\pm s_2$.
As summarized in Appendix,  the result is that one may effectively set $\Xi$ inside the 
proper-time integral as
\begin{align}
\Xi=&\left( 1+ \frac{\partial^2 B}{(eB)^2}\mathcal{C}_2 +\frac{(\partial B)^2}{(eB)^3}\mathcal{C}_3 \right),
\label{Xi=}
\end{align}
where $\mathcal{C}_2$ and $\mathcal{C}_3$ are certain coefficients obtained as polynomials in
two variables $\xi=-iw=eBT$ and $Y=\cot \xi$. 

Using \eqref{U=UcXi} and \eqref{Xi=} in \eqref{U=int}, one has the first two terms of the derivative expansions
\begin{align}
V&=
i\int_{0}^{\infty} \frac{dT}{T}\int d^2 \bar{x}\ e^{- i(\kappa  eB -  m^2 )T} \mathcal{N} \langle \bar{x}| U_c  |\bar{x}\rangle \cdot \Xi \nonumber \\
&= - \int d^2 x \left(\frac{eB}{4\pi }\right)^{3/2} 
\left[ {\alpha} +\frac{(\partial_j B)(\partial_j B)}{eB^3} {\beta}\right],
\label{V=i}
\end{align}
with the coefficients evaluated at $s=-\frac{1}{2}$ of
\begin{align}
 {\alpha}=&\sqrt{\frac{\pi}{2}}\mathcal{I}_{s}^\nu \left( 1 \right)
=2^{1-s}\Gamma(s)\zeta\left(s,\frac{1+\nu }{2}\right),          \label{alpha=sqrt}\\
 {\beta}=&
\sqrt{\frac{\pi}{2}}e^{-2}\mathcal{I}_{s}^\nu \left(\frac{e\mathcal{C}_2}{2}+ \mathcal{C}_3 \right),
\label{beta=sqrt}
\end{align}
where \eqref{alpha=sqrt} is equivalent to \eqref{V_c=}.
Note that both \eqref{alpha=sqrt} and \eqref{beta=sqrt} are order $e^0$.
Here we have introduced a linear operator 
\begin{align}
\mathcal{I}_{s'}^{\nu} (\mathcal{P})=\sqrt{\frac{2}{\pi}}\int_0^\infty dw \frac{e^{-  \nu w}w^{s'-1}}{\text{sh}\ w}\mathcal{P},
\label{Is'}
\end{align}
acting on the vector space spanned by polynomials $\mathcal{P}=\mathcal{P}(\xi, Y)$.
In Appendix, we systematically give the analytic continuations of
the proper-time integrals in the exponent $s$
using $\mathcal{I}_{s'}\equiv\mathcal{I}_{s'}^{\nu=1}$.
A useful basis for $\mathcal{P}$ is
\begin{align}
R_p=\text{sh}\ w \left(\frac{d}{dw}\right)^p \frac{w}{ \text{sh}\ w}.
\label{Rp}
\end{align}
The combination $\mathcal {S}=\frac{e}{2}\mathcal{C}_2+ \mathcal{C}_3$ in the argument in \eqref{beta=sqrt}
 comes from \eqref{Xi=} with the relation 
$\int d^2x (eB)^{-\frac{1}{2}} (\partial^2 B)=\frac{e}{2}\int d^2x ~(eB)^{-\frac{3}{2}} (\partial B)^2$,
which follows from an integration by parts assuming the surface term vanishes.
As shown in Appendix, this reads
\begin{align}
\frac{\mathcal{S}}{e^{2}}=\frac{w}{8} (R_3+R_1) + \frac{\kappa}{2}  \left[ \frac{R_1}{2} - w R_2\right] + \frac{\kappa^2 }{4} w R_1.
\label{S=}
\end{align}
Naturally, this reduces to the result for the 2+1d scalar QED \cite{cangemi1995derivative, gusynin1999derivative}
when $\kappa=0$ is taken.

We now specialize to the soliton-magnon system at low temperatures, where
we have $(m,\kappa)=(0,1)$ with the parameters $(s, \nu)=(-\frac{1}{2}, 1)$ 
of the physical interest as in \eqref{nu=}. 
The first coefficient in \eqref{alpha=sqrt} is obtained as 
\begin{align}
 {\alpha}=&2^{3/2}\Gamma\left(-\frac{1}{2}\right)\zeta\left(-\frac{1}{2}\right) 
=\sqrt{\frac{2}{\pi }} \zeta \left(\frac{3}{2}\right),
\label{alpha=zeta}
\end{align}
with the numerical value $\alpha\approx 2.08437$.
For the second coefficient in \eqref{beta=sqrt}, 
the polynomial  in \eqref{S=} leads to
\begin{align}
\beta
=&\frac{8 \pi ^2 \zeta \left(\frac{1}{2}\right)+18 \pi  \zeta \left(\frac{3}{2}\right)-15 \zeta \left(\frac{5}{2}\right)}{32 \sqrt{2} \pi ^{3/2}},
\label{beta=zeta}
\end{align}
with the numerical value $\beta\approx 0.048807$.
One has then the following small positive ratio
\begin{align}
\frac{ {\beta}}{ {\alpha}}\sim 0.02341\cdots~~. 
\end{align}
The derivative correction due to ${\beta}$ is shown to be about $9\%$
of the whole quantum energy corrections in Section \ref{section:proportional}.
Due to the remarkable property of the solitons satisfying the Bogomol'nyi relations, this turns out to be true
for any multi-soliton configurations. 
This suggests a special feature of the derivative expansion
that no additional small parameters be introduced by the details of the configurations; 
the convergence of it is of the same degree regardless of what configurations is considered.

Note that one has $\nu=\kappa$ for $m^2=0$ in \eqref{nu=}, which is expected for the magnon.
The result \eqref{alpha=sqrt} for the coefficient $\alpha$ implies
that the Casimir energy may change its sign as one varies 
$\kappa=0$ (the scalar QED) to $\kappa=1$ (the magnon-soliton system) since 
$\zeta(s,1)/\zeta\left(s,\frac{1}{2}\right)<0$ at $s=-\frac{1}{2}$.
Consequently, this suggests that the contribution to the quantum interaction
from $eA_\mu$ and that from $\kappa eB$, in a sense, compete with each other in \eqref{quadratic_L}. 
This point can be checked more explicitly in another point of view, the spectral zeta function as follows.

 \subsection{Spectral zeta function and the generalized Landau levels}
\label{section:spectral}


The purpose here is to compute the effective potential 
for the constant magnetic field using the spectral property for general values of $\kappa$,
which serves as an independent check of the first coefficient $\alpha$ in 
\eqref{alpha=sqrt} and \eqref{alpha=zeta}.
This also allows us to clearly show that 
the effective potential changes its sign exactly one time between $0<\kappa<1$.

A particularly useful view for the effective action comes from the spectral zeta function by Hawking
\cite{hawking1977zeta} and from related ideas for the QED by Dittrich \cite{dittrich1976one}.
This is a way to define 
a functional determinant in the form  
\begin{align}
-\mathrm{tr} \ln K =  \zeta_K'(0) + \cdots,
\label{zetaregularization}
\end{align}
where 
$K$ is some differential operator that governs the quadratic scalar fluctuations
by $\mathcal{L}=\varphi^* K \varphi$
such as that in \eqref{diracsquared};
``$\cdots$" includes some offset depending on the renormalization scale, which is unimportant below.
Let us denote the eigenvalues of $K$ by $\lambda_{\tilde{n}}$.
The spectral zeta function is then given by
\begin{align}
\zeta_K(s)= \sum_{\tilde{n}} \lambda_{\tilde{n}}^{-s}.
\label{zeta_K}
\end{align}

This view is particularly suitable for analyzing the problem in various dimensions, 
and works handily with the Euclidean metric.
As usual $K$ is to be associated with $i\varepsilon$ to make the path-integral convergent, 
corresponding to the definite sense $p_0 =i p_0^E$ of the Wick rotation of the integration path without hitting any poles.
To vary dimensions, one also needs to generalize the last term in \eqref{diracsquared}
to $\frac{e}{2}\sigma^{\mu\nu} F_{\mu\nu}$ with  
 $\sigma^{\mu\nu}=\frac{i}{2}[\gamma^{\mu},\gamma^{\nu}]$ being the commutator of the gamma matrices.
For simplicity, we start with the squared Dirac operator $K=-\slashed{D}^2$ for $d$-dimensional Euclidean massless QED in a uniform magnetic field.

Note now that the label $\tilde{n}$ of $\lambda_{\tilde{n}}$ is, in general, a set of both
discrete (e.g. $n$ in \eqref{landaulevel}) and continuous (e.g. momentum $p_z$) labels. 
Therefore, one has schematically $\sum_{\tilde{n}}\bullet=\widehat{\sum} \widehat{\int}\bullet$,
where $\widehat{\sum}$ and $\widehat{\int}$ are respectively a certain summation and integrals specified in the following.
One has the eigenvalues \cite{brezin1970pair}
\footnote{These non-negative eigenvalues contain those of $-\mathcal{X}$ with $\mathcal{X}$ in \eqref{ln_identity}; the zero eigenvalues are automatically removed by analytic continuation of the zeta function. The sign convention in \eqref{TV=minus_i} from \eqref{Z=} does not change the result.}
 of the form
 $\lambda_{\tilde{n}}=\bm{p}^2 +2m E_{n, \pm}$, which reads
\begin{align}
\lambda_{\tilde{n}}^E=(p_{0}^E)^2+ p_1^2+\cdots +p_{d-3}^2 +2eB \left[n+ \frac{1\pm \kappa}{2} \right].
\label{lambda=}
\end{align}
 To count the multiplicities of the eigenvalues, consider the QED in a box of volume $L^d$.
 Each Landau level has the multiplicity given by the number of the flux quanta
 $BL^2/\left(2\pi /e \right)$ on the slab of area $L^2$.
 Then, the summation $\sum_{\tilde{n}}$ may be decomposed into
 a discrete sum $\widehat{\sum}\equiv\left(\frac{eBL^2}{2\pi}\right) \sum_{n, \pm}$ and
momentum integrations transverse to the cyclotron plane
$\widehat{\int}\equiv\left(\frac{L}{2\pi}\right)^{d-2}\int dp_0^E\cdots dp_{d-3}$. 
Starting by setting $A^2\equiv \lambda_{\tilde{n}}^E $, each of such $d-2$ integrations may be performed iteratively with
\begin{align}
\int_{-\infty}^{\infty} dp~ (A^2 + p^2)^{-s} = (A^2)^{-s+\frac12}\frac{\Gamma\left(\frac{1}{2}\right) \Gamma\left(s-\frac{1}{2}\right)}{\Gamma\left(s\right)},
\end{align}
resulting in an increment of the exponent $s \mapsto s-\frac12$ until $A^2$ reduces to $2eB[n+(1\pm \kappa)/2]$.
For the soliton-magnon system, we only need {the minus branch} of \eqref{lambda=};
see the last paragraph of Section \ref{section:magnon-soliton}.
We denote the corresponding operator by $K_-$.
The remaining sum over $n$ yields
\begin{align}
\frac{\zeta_{K_-}(s)}{L^d} =  \frac{(2eB)^{\frac{d}{2}-s}\zeta\left(s-\frac{d-2}{2} , \frac{1- \kappa}{2} \right)
}{2(2\pi)^{d-1}}
\frac{\pi^{\frac{d-2}{2}} \Gamma\left(s-\frac{d-2}{2}\right)}{\Gamma\left(s\right)},
\label{zeta_s}
\end{align}
where the
Hurwitz zeta function $\zeta(s, x)$ is defined as an analytic continuation of the series $\sum_{n=0}^\infty (n+x)^{-s}$.
Since the only singular factor around $s=0$ is $\Gamma(s)=s^{-1}+\mathcal{O}(1)$, it is easy to evaluate the first derivative
 in \eqref{zetaregularization}. 
By setting the physical values $(d,\kappa)=(3,1)$, this yields
\begin{align}
\zeta_{K_-}'(0)=-\mathrm{tr} \ln K_- \sim \frac{(eB)^{\frac{3}{2}}\zeta\left(-\frac{1}{2},0\right)\Gamma\left(-\frac{1}{2}\right)}{2\sqrt{2}\pi^{3/2}} L^3.
\label{trlnK}
\end{align}
Now the overall sign of the effective potential $V$ is the minus of \eqref{trlnK}; this may be seen by the Wick rotating back, where the Minkowski zeta function
$\zeta_{M, K_-}(s)=i\zeta_{K_-}(s)$ 
(see below \eqref{zeta_K} for the sense)
is naturally defined by replacing 
$\int_{-\infty}^{\infty} dp_0^E $ in $\zeta_{K_-}(s)$ by  $\int_{-\infty}^{\infty} dp_0$. 
Note that we have introduced the effective action in the form $V\propto i\ln Z = -i \mathrm{tr} \ln K_{-}$ in \eqref{TV=minus_i}. This leads to $V \propto i\zeta_{M, K_-}'(0) = - \zeta_{K_-}'(0)$.
Therefore, the first coefficient $\alpha$ in \eqref{alpha=zeta} 
in the derivative expansion \eqref{V=i} is exactly reproduced.

It is remarkable in \eqref{zeta_s} that the $\kappa$-dependence is solely absorbed into the argument of the Hurwitz zeta function.
Since $\zeta\left(s, 1\right)=\zeta\left(s, 0\right)=\zeta(s)$, the choice within $\kappa=\pm 1$ 
does not change the result at this level
\footnote{For this reason, the result \eqref{alpha=zeta} equals,
up to fermion loop factor $(-1)$, 
that of the spinor QED \cite{cangemi1995derivative, gusynin1999derivative}.}.
Thus, we have a spectral flow $n+\frac12 \to n$
in the eigenvalues $E_{n, -}$ from the scalar QED ($\kappa=0$) to the soliton-magnon system ($\kappa=1$), which changes the effective action continuously.

Note that to study the massless scalar QED\cite{cangemi1995derivative, gusynin1999derivative},
one can simply replace the factor $\zeta\left(-\frac{1}{2}, 0\right)$ in \eqref{trlnK}
by $\zeta\left(-\frac{1}{2},\frac{1}{2}\right)$.
These factors have opposite signs since they are related by the following identity:
\begin{align}
\zeta\left(\tilde{s}, \frac{1}{2}\right) = (2^{\tilde{s}}-1) \zeta\left(\tilde{s} \right).
\label{zetas}
\end{align}
 Since the argument is $\tilde{s}=-(d-2)/2$ as seen from \eqref{zeta_s} 
as $s \to 0$, this sign flip occurs 
whenever for $d>2$ as one 
deforms the theory \eqref{quadratic_L} from $\kappa=0$ to $\kappa=1$.

The upshot is an inclusion of $\kappa eB |\varphi|^2$ (denoted below by the subscript ``$P$": the Pauli term) 
in \eqref{quadratic_L} 
on top of $\left|\mathcal{D}_{\mu}\varphi\right|^2$ in the scalar QED changes the sign of the Casimir energy.  
This is, in a sense, analogous to the interplay between 
the Landau diamagnetism ($\chi_D$) and the Pauli paramagnetism ($\chi_P$) 
in the magnetic susceptibility of free electron gas\cite{landau1930diamagnetismus},
where $\chi_P$ turns out to be dominant due to the celebrated ratio $\chi_D : \chi_P = 1:-3$.
In fact, the relation between the Hurwitz zeta functions \eqref{zetas} reads 
\begin{align}
{V}^{\mathrm{QED}}_{``D"}=(2^{-1/2}-1){V}^{\mathrm{soliton}}_{``D+P"},
\end{align}
where
${V}^{\mathrm{QED}}_{``D"}$ and 
${V}^{\mathrm{soliton}}_{``D+P"}$
respectively 
stand for the effective potential (the Casimir energy) in the system \eqref{quadratic_L}
with $\kappa=0$ and $\kappa=1$
in the presence of the constant magnetic field $B$.

As a technical remark, one may wonder whether with some 
further refinement of the computation in this subsection by using the heat kernel approach, for instance, one could reproduce the next-order coefficient $\beta$ in \eqref{beta=zeta} as well.
A known example in QED ($\kappa=0$) suggests, however, that the heat kernel expansion 
\footnote{See Sec. 4.3 in Ref. \onlinecite{dittrich2000probing} for the erratum to 
the paper using the heat kernel expansion\cite{dittrich1997flavor}.}
works for the weak field case $|eB|\ll m^2$,
but not in the massless limit $m\to 0$.
Currently, the worldline formalism adopted in Section \ref{section:derivative}
is therefore the only method that allows us to obtain the coefficient \eqref{beta=zeta}
for the derivative term.




\section{Quantum Interactions of Solitons} \label{section:quantum}

\subsection{One soliton and a concentric soliton rings for an arbitrary $q$} 
The general formula \eqref{V=i} can be used to compute the quantum correction to the energy in the presence of solitons.
Let us take the simplest case, where $q$ of the solitons of size $\lambda$ forms a concentric ring 
configurations, for which the magnon scattering off the soliton have been studied \cite{rodriguez1989quantized, walliser2000casimir, ivanov2007quantum}. 
We therefore consider the magnetic field \eqref{B_nsoliton}
as a function of the distance $r$ from the center:
\begin{align}
B (r)= \frac{ 4q^2 \lambda^{2q} r^{2q-2}}{\left(  \lambda^{2q}+r^{2q} \right)^2}.
\end{align}
If this is integrated over the whole plane as in \eqref{topological8piq}, by construction, 
the topological invariant, namely, the wrapping number times $\mathrm{Vol}(S^2)=4\pi$   
should appear
\begin{align}
\int_{0}^{\infty} 2\pi r dr\ B(r) =4\pi q.
\end{align}
Importantly, due to \eqref{topological4piq} and \eqref{A_tildeA}, 
this is exactly the double of the classical energy bound in \eqref{Bogomolnyi}, 
which is scale-invariant and has no $\lambda$-dependence. 

The quantum energy correction \eqref{V=i}, on the other hand, may be evaluated from the following basic blocks  
\begin{align}
\int_{0}^{\infty} 2\pi r dr\ B(r)^{{3}/{2}} =&\frac{f(q)}{\lambda },                     \label{fq1}\\
\int_{0}^{\infty} 2\pi r dr\ \frac{(\partial_r B(r))^2}{B(r)^{{3}/{2}}} =&
\frac{4f(q)}{\lambda },   \label{fq2}
\end{align}
where the non-linear dependence on $q$ is given by
\begin{align}
f(q)=\frac{\pi ^2 \left(q^2-1\right)}{\cos \left({\pi }/{2 q}\right)}, 
\label{fq=}
\end{align}
which has a simple 1-soliton value $\lim_{q\to 1}f(q)=4\pi$ and 
the asymptotic behavior $f(q)= \pi^2 q^2 +\frac{\pi ^4}{8}-\pi ^2 +\mathcal{O}(q^{-2})$
as $q\to \infty$.
The effective potential \eqref{V=i} in the case of one soliton becomes a function of size
\begin{align}
V_q(\lambda)=&-\sigma \frac{f(q)}{\lambda }, 
\label{Vq=}
 \end{align}
with a positive coefficient $\sigma$, via \eqref{alpha=zeta} and \eqref{beta=zeta}, yielding
\begin{align}
\sigma=\frac{{\alpha}+ 4 \beta}{\left(4\pi\right)^{\frac{3}{2}} }=\frac{8 \pi ^2 \zeta \left(\frac{1}{2}\right)+34 \pi  \zeta \left(\frac{3}{2}\right)-15 \zeta \left(\frac{5}{2}\right)}{64 \sqrt{2} \pi ^3},
\label{sigma=} 
\end{align}
with the numerical value $\sigma\approx 0.05117$. Without the derivative correction,
the result decreases by $9\%$ and $\sigma=\zeta \left(\frac{3}{2}\right)/(4 \sqrt{2} \pi ^2)\approx 0.04679$.
The classically stable soliton, which may have any size $\lambda$, now acquires the Casimir energy that depends on $\lambda$. Such phenomenon is also referred to as the breaking of the conformal invariance \cite{rodriguez1989quantized, walliser2000casimir, ivanov2007quantum}
in the original $O(3)$ model.

Using the general result \eqref{fq=} in \eqref{Vq=} with \eqref{sigma=}, 
we obtain the Casimir energy for $q=1$ and $q=2$ cases, yielding
\begin{align}
V_1\approx -\frac{0.643}{\lambda},\qquad V_2\approx -\frac{2.14}{\lambda}.
\label{V1V2}
\end{align}
This could be compared with the result computed using a generalization of the DHN formula \cite{dashen1974nonperturbative, rajaraman1982solitons}
by evaluating the scattering phase shifts numerically\cite{walliser2000casimir}. 
The corresponding result
\footnote{Ref. \onlinecite{walliser2000casimir} is the only choice  
since the result\cite{rodriguez1989quantized} is corrected by 
the numerics\cite{walliser2000casimir}; 
Ref. \onlinecite{ivanov2007quantum} deals with ferromagnets only.
}
 (i.e. Eq. (32) of Ref. \onlinecite{walliser2000casimir}) reads
\begin{align}
E^{\mathrm{cas}}_1\approx -\frac{0.5}{\lambda},\qquad E^{\mathrm{cas}}_2\approx -\frac{1.7}{\lambda},
\label{E1E2}
\end{align}
by correctly identifying the parameter $R_n$ to our size parameter $\lambda$ in \eqref{B_nsoliton}.
According to \eqref{V1V2} and \eqref{E1E2}, the sign and the qualitative $1/\lambda$-behavior 
agrees with each other. 
The relative ratios ${V_2}/{V_1}= \frac{3 \pi }{2 \sqrt{2}}\approx 3.33$
and $E^{\mathrm{cas}}_2/E^{\mathrm{cas}}_1\approx 3.4$ also agrees, while the normalization used in Ref. \onlinecite{walliser2000casimir} seems to differ by factor of $\sqrt{2}$ from that of us due to the convention used in the spin stiffness. 

Instead of dwelling further on the numerical comparison, we here multiply \eqref{V1V2} by $\hbar c_s$
and restore $\hbar$ and the (bare) spin wave velocity $c_s$ given below \eqref{S=SB}. 
This yields an energy scale
$V_1(\lambda)\sim -1.8 JS a/\lambda$.
According to \eqref{Bogomolnyi} and \eqref{S=SB},
the classical energy gap $E_g$ is given by $E_g=4\pi \hbar\rho_s\sim 4\pi JS^2$ 
between the nearest-neighbor topological sectors.
It is also possible to observe that the quantum correction $V_1$ never exceeds the gap $E_g$
 unless $\lambda< a$ for the atomic spin $S\geq \frac{1}{2}$.
As the soliton size $\lambda$ should at least span several multiples of $a$, one usually expects $|V_1| \ll E_g$.
On the other hand, the smoothness assumption used in \eqref{S=SB} 
can not be expected to hold all the way to the regime $\lambda\lesssim a$.

Our result as well as the previous studies\cite{walliser2000casimir,ivanov2007quantum} suggests, in particular, that single-soliton of size $\lambda\gg a$ becomes unstable and starts to shrink. 
As discussed in ferromagnets \cite{ivanov2007quantum}, 
if the soliton shrinks eventually to the lattice spacing $a$, it may evaporate and a quantum increment (or a reduction)
of the topological charge e.g. $q=n \to q=n\pm1$
could happen. 
For an isolated soliton, this process should look like a hedgehog configuration \cite{haldane1988monopole}
of spin directions in 2+1d space-time picture. 
We note, however, that our assumption of the smooth configuration used in \eqref{S=SB} 
should break down before reaching $\lambda\sim a$ as argued above.

\subsection{The Bogomol'nyi relation implies a proportionality relation between the first two terms}\label{section:proportional}
To obtain the Casimir energy for multi-solitons, 
it turns out to be much simpler computing $V$ in \eqref{V=i} than 
solving a magnon scattering problem, which is necessary in the DHN framework, 
each time for each given a soliton-configuration. 
By evaluating the derivative expansion \eqref{V=i} for various magnetic field configuration 
induced from the solitons, one should notice that 
the first two terms, $\int d^2x\ B^{3/2}$ and $\int d^2x\  (B^{-3/2}\partial_j B\partial_j B)$,
are proportional to each other.
The simplest example of this phenomena 
has appeared already in  \eqref{fq1} and \eqref{fq2}, where the common dependence $f(q)$ 
appears in the numerators of the both terms.
Actually, such a remarkable relation holds for 
any configurations of multi-solitons as long as they saturate the Bogomol'nyi inequality.
This can be shown generally in a short paragraph as follows.

By virtue of the CR relation $0=\bar{\partial} W=\partial \bar{W}$ in \eqref{cauchyriemann},
we find the magnetic field \eqref{BWW} satisfies a non-linear differential equation 
\begin{align}
B\bar{\partial} \partial B =
\bar{\partial}B \partial B -2B^3,
\label{Liouville}
\end{align}
when the Bogomol'nyi relation holds.
One can then use this local relation to show the global non-trivial relation
\begin{align}
\int d^2x\ \frac{\bar{\partial}B \partial B}{B^{{3}/{2}}} =
4\int d^2x\ B^{{3}/{2}}. 
\label{BPSmagic}
\end{align}
by an integration by parts, in which we assume the surface term vanishes for generic cases. This completes the proof of the claim.

Consequently,  one has to only compute the integral
\begin{align}
V=-\sigma \int d^2x~ B(x)^{3/2}
\label{V=sigma}
\end{align}
with the single coefficient $\sigma$ in \eqref{sigma=}.
This formula takes account the fact that the second term with derivatives in \eqref{V=i}  always gives $\sim 9.366\%$-corrections to the first term without derivatives
for any BPS-soliton configurations. 

Some remarks on the higher-order derivative terms are in order.
In view of the derivative expansion, we have shown that the BPS soliton configuration has a rather remarkable property that the derivative correction is proportional to the main term. 
It is then tempting to consider the same is true for the terms with more derivatives, which means the convergence of the expansions after resummation is not controlled by the slow variation of $B$ but by the coefficients themselves. 
Although a standard dimensional analysis restricts the forms of the higher terms,
the coefficients of the terms with more than two derivatives have never been computed  
even in the best studied case of the QED for generic field configurations.
Working out further coefficients is thus not simple task; it would be still interesting to pursue a possible relation between this general phenomena and the exactly solvable, inhomogeneous field configurations of the QED \cite{cangemi1995effective}, where the derivative expansions is computed for an arbitrary higher order in the derivatives. 

Geometrically, the effective magnetic field $B$ may be interpreted as a conformal factor \cite{dubrovin1992modern} for the metric induced by the meromorphic function $W(z)$ (satisfying $\bar{\partial}W=0$ except at simple poles).
Indeed, by substituting $B=e^{\phi}$ in \eqref{Liouville}, one gets the Liouville equation
$\bar{\partial}\partial\phi=-2e^{\phi}$ for a positive constant curvature surface\footnote{The authors are grateful to Paul Wiegmann for suggesting this simple substitution in \eqref{Liouville}.},
whose general solution is given by \eqref{BWW}.
In this view, $B$ is proportional to the area element, 
and \eqref{topological8piq} is the Gauss-Bonnet formula.
It is then an interesting open question whether one can give a meaning to the quantum invariants constructed from the global relations like \eqref{BPSmagic} or its higher generalizations, as they essentially involve $\sqrt{B}$.

\begin{figure}
\includegraphics[width=8cm]{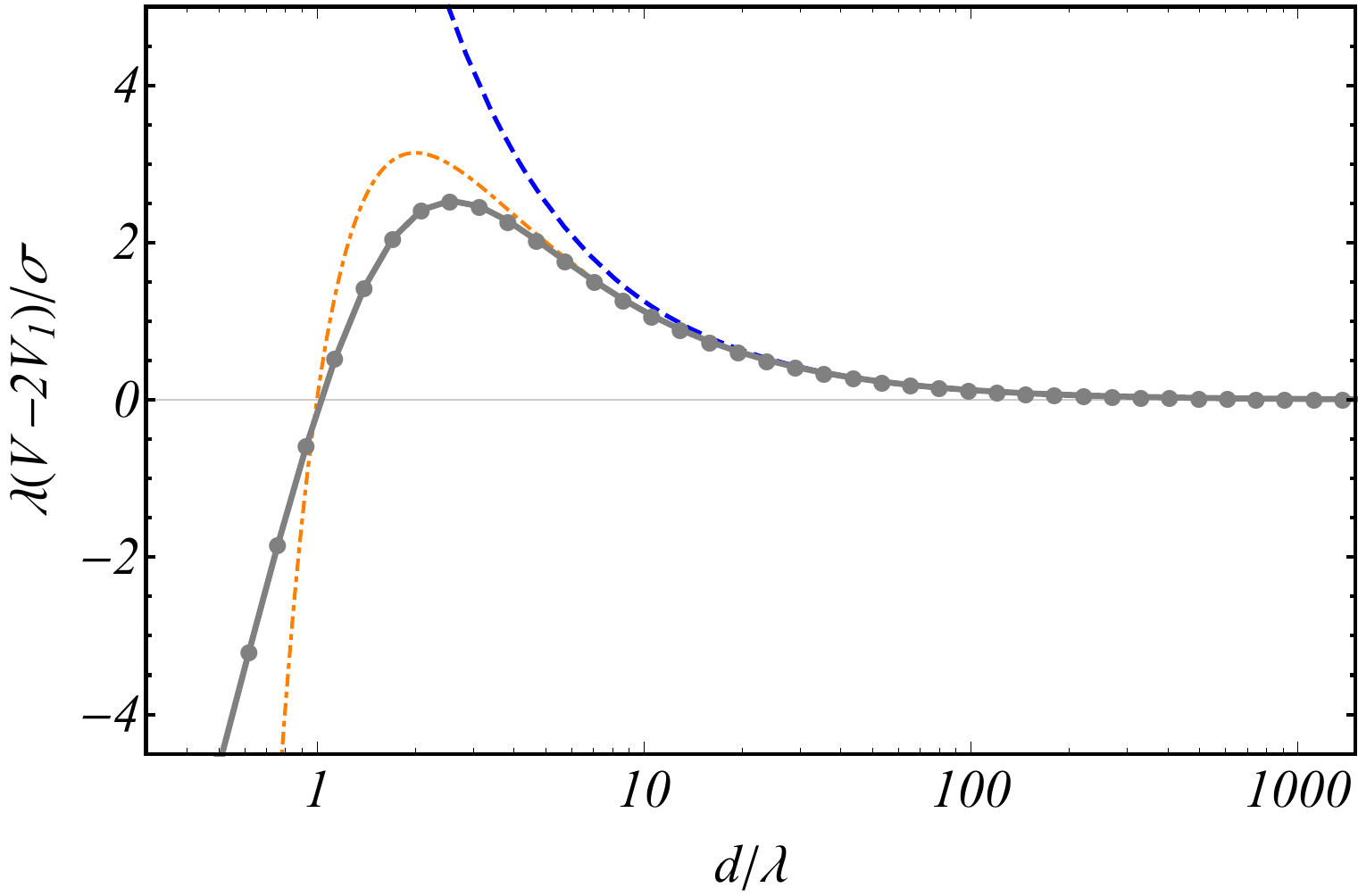} 
\caption{\label{fig:V_2soliton}
The quantum interaction potential $V(\lambda, d)$ for two static solitons of size $\lambda$  
subtracted by the self-energy $2V_1(\lambda)$
with a separation $2d$ (solid with markers).
An attractive well of depth $\approx -16.7$ lies at $d=0$ for the relative orientation $\theta=\pi$.
The short-range interaction depends on $\theta$ (see FIG. \ref{fig:2solitons}),
while the long-range one \eqref{large_d} is universal.
The leading repulsive $1/d$-law (dashed) and the next order (dash-dotted). 
} 
\end{figure}

\subsection{The quantum interaction potential for 2-solitons } 

The Casimir energy for multi-solitons had remained more non-trivial to be evaluated. 
The quantum 2-soliton potential was first evaluated in the DHN framework with the Born approximation \cite{rodriguez1989quantized},
the use of which was later criticized by the same framework but with numerics \cite{walliser2000casimir}. 
Ref. \onlinecite{walliser2000casimir} was, however, not able to discuss more than 1-soliton configurations. 
Our derivative expansion \eqref{V=i} could offer more flexible method to work with multi-solitons 
as long as they satisfy the Bogomol'nyi relation.

For 2-solitons of the same size $\lambda$ with a distance $2d$ apart, 
one can apply \eqref{V=sigma} for the magnetic field obtained in \eqref{B_2soliton}.
In a polar coordinate, this leads to
\begin{align}
V(\lambda,d)=-\sigma\int_0^\infty rdr \int_0^{2\pi}d\theta ~\frac{64r^3\ell^6 }{(a+b \cos 2 \theta)^3},
\end{align}
with $a=\ell^4+d^4+r^4$, $b= 2 d^2 r^2$, and $\ell^2=\lambda (2 d+\lambda )$.
One can start with performing either $r$- or $\theta$-integral analytically.
Here the latter is done by the formula,
\begin{align}
\int_0^{2\pi}\frac{d\theta}{(a+b \cos 2 \theta)^{n+1}}= \frac{2\pi P_n\left( \frac{a}{\sqrt{a^2-b^2}} \right)}{\left(a^2-b^2\right)^{\frac{n+1}{2}}},  
\label{general_expression}
\end{align}
for $n=2$ with the Legendre polynomial $P_2(x)=\frac{3x^2-1}{2}$, which results in a general expression
\begin{align}
V(\lambda,d)=-\sigma \int_0^\infty dr~
\frac{64\pi r^4\ell^6 \left(2 a^2+b^2\right)}{ \left(a^2-b^2\right)^{5/2}}.
\end{align}
For large $d$, it is easy to notice the integrand is localized around the region $|r -d|=\mathcal{O}(\lambda)$
with a power-law tail $\sim |r -d|^{-5}$.
A careful expansion then leads to the asymptotic behavior
\begin{align}
V(\lambda,d)=-\sigma \frac{8\pi}{\lambda} \left[1- \frac{\lambda}{2d} + \frac{\lambda^2}{2d^2}   + \mathcal{O}\left( \frac{\lambda^3}{d^3} \right) \right],
\label{large_d}
\end{align}
as $d\to \infty$. The first term is naturally $2V_1(\lambda)$, where $V_1(\lambda)$ is the Casimir energy 
 \eqref{Vq=} for an isolated $q=1$ soliton obtained from $f(1)=4\pi$.
If this self-energy is subtracted, what remains is an interaction energy shown in FIG. \ref{fig:V_2soliton}.
According to \eqref{large_d}, 
this is a universal repulsive ${1}/{d}$-potential, which does not depend on the soliton size at the leading order.

On the other hand, one obtains the expansion
 \begin{align}
V(\lambda,d)=-\sigma \frac{3\sqrt{2}\pi^2}{\lambda} \left[1- \frac{d}{\lambda} + \frac{3d^2}{2\lambda^2}   
+ \mathcal{O}\left( \frac{d^3}{\lambda^3} \right) \right].
\label{short_d}
\end{align}
for small $d\ll \lambda$. Note that here we restrict the analysis to the case $\theta=\pi$ in FIG. \ref{fig:2solitons}-(c).
The higher-order terms are alternating and not useful to grasp a global profile 
at least without some elaborate resummation techniques. 
The first term is the energy $V_2(\lambda)$ of the overlapped $q=2$ solitons with $f(2)=3 \sqrt{2} \pi ^2$.
Since $f(2) - 2 f(1)>0$, the interaction energy is negative at $d=0$ suggesting the potential
has an attractive well of $V_{\mathrm{int}}=-\sigma\pi(3\sqrt{2}\pi-8)/\lambda\approx
- 16.7\sigma /\lambda$. 
In general, the convexity of \eqref{fq=} suggests that the static solitons localized in a finite region 
($d\lesssim \lambda$) 
become more stable when they overlap
despite their repulsive long-ranged interaction for $d\gg \lambda$.
The two limits of $V(\lambda, d)$ are connected smoothly with one maximum around $d\approx 3\lambda$. 
This value corresponds to the green-colored plane in Fig. \ref{fig:distance}, below which the interaction may be attractive.

Among other things, a direct measurement of the kinetic energy gained by the universal repulsive force in \eqref{large_d}, which is independent of the size and of the orientation, using recently developing ideas in the skyrmion racetrack memory on a nanostripe would be interesting. Below we give a sketch of the experimental setting although it is too crude to perform as it is. For instance, the last step of measuring tiny kinetic energy seems to be missing; most of the current studies assume the Dzyaloshinskii-Moriya interaction (DMI), which is absent from \eqref{S=SB}, to stabilize the skyrmions. Such problems may be, however, overcome in near future given that the skyrmion manipulation is rapidly developing field.

According to the micromagnetic simulation in Ref. \onlinecite{jin2016dynamics}, two antiferromagnetic skyrmions of the same size $\lambda\sim 20$ nm may be nucleated by injecting the spin-polarized pulses of $0.1$ ns in the middle of the antiferromagnetic nanostripe followed by a longer relaxation of a few nanoseconds. With tiny (or  without) DMI, this would bring these skyrmions into nearly the BPS form in the $q=2$ sector. Remarkably, our solitons (antiferromagnetic skyrmions in contrast to ferromagnetic ones) has a great advantage that they can be moved in a straight line without deflection by applying a current; this is due to the mutual cancellation of the Magnus forces from the two magnetic sublattices \cite{barker2016static, zhang2016antiferromagnetic, jin2016dynamics, smejkal2018topological}. The current is used to move the skyrmions, and then triangular notches of zero saturation magnetization\cite{jin2016dynamics} is used to pin both the skyrmion centers at the designated distance $2d$. After turning off the current, the kinetic energy due to the initial acceleration is measured for various $d\gtrsim 3\lambda$. We here note two relevant energy scales again by multiplying $\hbar c_s$: the leading interaction energy $V_{\mathrm{int}}(d) \sim 1.8JS a/d$ in \eqref{large_d}
and the energy barrier height $V_{\mathrm{max}}(\lambda) \sim 0.37JS a/\lambda$ between \eqref{large_d} and \eqref{short_d}.

Qualitatively, the result that the solitons has 
a repulsive long-range potential with an attractive well
agrees with the previous result \cite{rodriguez1989quantized}. 
The independence of the long-range potential on the relative orientation $\theta$ 
is also shared by both our work and by Ref. \onlinecite{rodriguez1989quantized}.
There are also relatively minor quantitative differences; 
our $1/d$-law differs from the $1/d^{2}$-law\cite{rodriguez1989quantized}, 
in which (accordingly by dimensional analysis) the potential depends on the sizes of the solitons 
at the leading order. 
Such a discrepancy is, however, not surprising;  
as pointed out in Ref. \onlinecite{ivanov2007quantum}, 
the Born approximation\cite{rodriguez1989quantized} may not correctly account
the AB-scattering phase because of the long-range nature of the topological-soliton gauge potential.
A possibility is that our worldline formulation yields the derivative expansion \eqref{V=i} 
around the exact solution for the constant field \cite{nambu1950use, schwinger1951gauge}
and may be capable of accounting the long-range quantum phase more appropriately.

\section{Conclusion}\label{section:conclusion}

We present a new quantitative approach to the quantum interaction 
of the topological solitons \cite{belavin1975metastable} 
(magnetic skyrmions) in antiferromagnets based on the worldline formulation\cite{feynman1950mathematical,strassler1992field,schubert2001perturbative}  
of the QED in an external field\cite{nambu1950use}. 
The Casimir energy for classically non-interacting solitons that saturate the Bogomol'nyi bound
arises in the same way as the Maxwell Lagrangian gets the non-linear interactions in the QED\cite{heisenberg2006consequences};
the solitons and the magnons respectively play the role of the external magnetic field 
and the massless scalar particles in QED. 

We start with the standard Lagrangian \eqref{S=SB} describing the 
magnon excitations 
in antiferromagnets in 2+1d.
This trivially contains the 2d classical non-linear sigma model \eqref{H_O3}, which allows the topological soliton configuration.
The crucial observation is that the energy \eqref{quadraticF} quadratically expanded in 
the fluctuation around the solitons essentially relates to the Pauli Hamiltonian \eqref{pauliNonrela},
which is the non-relativistic limit of the QED.
Based on these fact, we are led to derive the 2+1d relativistic field theory \eqref{quadratic_L}
that governs the magnon-soliton coupling.

To obtain the quantum effective potential,
the path-integral is performed over the magnons in the two distinct views, namely, 
the worldline formulation and the spectral zeta function\cite{hawking1977zeta,dittrich1976one}  .
In the worldline formulation, we effectively map the field theory \eqref{quadratic_L} to the non-relativistic
Lagrangian \eqref{L=} in the first quantization resulting in the derivative expansion \eqref{V=i}.
In the spectral zeta function,  we similarly map \eqref{quadratic_L} to 
the non-relativistic Hamiltonian \eqref{pauliNonrela}, which enables us to check the first term in \eqref{V=i}. 
Both maps are understood from the use of the proper-time evolution along the worldline, 
namely in our case, 
from the viewpoint from the single magnon itself; a non-relativistic quantum mechanics
may serve as an efficient and non-perturbative description of the one-loop sector of a relativistic QFT. 

Extracting the quantum interaction of the solitons from \eqref{V=i} could be, in principle, as complicated as solving the scattering problem necessary in the conventional DHN framework \cite{dashen1974nonperturbative, rajaraman1982solitons} for the soliton quantization
\cite{rodriguez1989quantized, ivanov2007quantum, walliser2000casimir}. 
To our surprise, however, the analysis is significantly simplified by the 
remarkable identity \eqref{BPSmagic} due to
the Bogomol'nyi relation.

This enables us, for instance, to work out the instability of the concentric soliton ring configuration
\eqref{Vq=} or the quantum effective potential \eqref{large_d} between two solitons.
We have shown, in particular, that
the two solitons of size $\lambda$ with a separation $d$ may attract with each other when they overlap $d\lesssim \lambda$, but they repel by the universal $1/d$-law when they are far apart $d\gg \lambda$.
Although the qualitative behaviors are similar to those\cite{rodriguez1989quantized} 
obtained in the DHN framework, our approach seems to be much more flexible to handle 
various multi-soliton configurations as long as they saturates the Bogomol'nyi bound.

It would be interesting to find whether some non-abelian generalization of the ``gauge theory in the external field" 
picture
works for quantizing solitons in a model with higher symmetries\cite{ivanov2008pairing, ueda2016quantum} such as the $SU(3)$ relevant for the spin-1 cold atom whose continuum physics is described by the $\mathrm{CP}^2$ model. 
A few extensions of this work would be the analysis of the interaction between the merons\cite{gross1978meron}, 
the three-body interaction of the solitons, or the Casimir energy for a special soliton-of-solitons configuration,
where the corresponding magnetic field itself becomes the soliton profile 
being localized in a strip of a given width\cite{cangemi1995effective},
which would be relevant for the stability of the skyrmion lattices against the quantum fluctuations. 

\begin{acknowledgments}
The authors thank Yutaka Akagi, Koji Hashimoto, Shinobu Hikami, Masaru Hongo, Karlo Penc, Nic Shannon, Hidehiko Shimada, Tokuro Shimokawa, and Paul Wiegmann for useful comments.
The work of H. S. is supported in part by the Okinawa Institute of Science and Technology 
 Graduate University, and by JSPS KAKENHI Grant No. 16K05491 and No. 18K13472.
The work of K. T. is supported by JSPS KAKENHI Grant No. 26400385.
\end{acknowledgments}%

\appendix*
\begin{table}
\caption{\label{tab:1}The form of the polynomials $R_p$ and 
their values when acted by the linear operator $\mathcal{I}_{s'}(\cdot)$ with $s'=\pm\frac{1}{2}$.
A simplified notation $\zeta_x\equiv \zeta(x)$ is used in the row $p=3$.}
\begin{ruledtabular}
\begin{tabular}{c|r|r|} 
$p$&  $R_p=\left[\text{sh}\ w\left(\frac {d}{dw} \right)^p \frac {w} {\text{sh}\  w}\right]_{w=i\xi}$ & 
$\mathcal{I}_{\frac{1}{2}}(R_p)=\mathcal{I}_{-\frac{1}{2}}(w R_p)$
\\
\hline
$0$ & $i \xi$   & $\frac{1}{2} \zeta \left(\frac{3}{2}\right)$ \\
 $1$ & $-\xi  Y +1$  & $\zeta \left(\frac{1}{2}\right)  {+}\frac{1}{2} \zeta \left(\frac{3}{2}\right)$\\
$2$ & $-i \left[2  \xi  Y^2 - 2Y  +\xi  
\right]$
& $-6$ $\zeta \left(-\frac{1}{2}\right)  {+}2 \zeta \left(\frac{1}{2}\right)+\frac{1}{2} \zeta \left(\frac{3}{2}\right)$\\
 $3$ & $6 \xi  Y^3 -6 Y^2 +5 \xi Y -3$  & $20 \zeta_{-\frac{3}{2}} {-}18 \zeta_{-\frac{1}{2}} +3 \zeta_{\frac{1}{2}} {+} \frac{1}{2} \zeta_{\frac{3}{2}}$\\
& & \\
\hline
\hline
$p$&  $R_p=\left[\text{sh}\ w\left(\frac {d}{dw} \right)^p \frac {w} {\text{sh}\  w}\right]_{w=i\xi}$ & 
$\mathcal{I}_{-\frac{1}{2}}(R_p)$
\\
\hline
 $1$ & $-\xi  Y +1$  & $-12 \zeta \left(-\frac{1}{2}\right)  {+} 2 \zeta \left(\frac{1}{2}\right)$\\
\end{tabular}
\end{ruledtabular}
\end{table}

\section{Worldline perturbations: anatomies for the derivative term coefficient $ {\beta}$}\label{integraltable}
The second order perturbation series $\Xi$ in \eqref{Xi} can be cast into the 
form \eqref{Xi=} by performing integrations over the proper-time along the magnon loop,
which yields \eqref{S=}. 
In this appendix, we outline somewhat technical computations leading to \eqref{S=}.
The longest part has been done\cite{cangemi1995derivative, gusynin1999derivative}
 as given in \eqref{A1A2=}, while the rests are relatively simple as shown below. 

At the first order in $L_1=L_{10}+L_{11}$ in \eqref{L1=}, one has,
\begin{align}
\label{A1}\mathcal{A}_1=&i\int_0^T ds~ \langle L_{10}(s) \rangle,\\
\label{B}\mathcal{B}=&i \int_0^T ds~ \langle L_{11}(s) \rangle.
\end{align}
At the second order, one has,
\begin{align}
\label{A2}\mathcal{A}_2=&-\frac{1}{2}\int_0^T\int_0^T ds_1ds_2~\langle L_{10}(s_1)L_{10}(s_2) \rangle,\\
\label{C}\mathcal{C}=&2\cdot \frac{-1}{2}\int_0^T\int_0^T ds_1ds_2~\langle L_{10}(s_1)L_{11}(s_2) \rangle,\\
\label{D}\mathcal{D}=&-\frac{1}{2}\int_0^T\int_0^T ds_1ds_2~\langle L_{11}(s_1)L_{11}(s_2) \rangle.
\end{align}
After dealing a bit with \eqref{A1}-\eqref{D} 
using $L_{10}$ and $L_{11}$ in \eqref{L10}-\eqref{L11}, one recognizes the following form,
\begin{align}
\mathcal{A}_1+\mathcal{B}  =  \frac{\partial^2 B}{(eB)^2}\ \mathcal{C}_2,\quad     
\mathcal{A}_2+\mathcal{C}+\mathcal{D} = \frac{(\partial B)^2}{(eB)^3}\ \mathcal{C}_3.
\label{A1B}
\end{align}
It is then handy to introduce the tilde coefficients by
$\mathcal{A}_1=\tilde{\mathcal{A}_1}{(\partial^2 B)}/{(eB)^2}$,
$\mathcal{A}_2=\tilde{\mathcal{A}_2}{(\partial B)^2}/{(eB)^3}$, and so on, so that
$\mathcal{C}_2=\tilde{\mathcal{A}_1}+\tilde{\mathcal{B}}$ and
$\mathcal{C}_3=\tilde{\mathcal{A}_2}+\tilde{\mathcal{C}}+\tilde{\mathcal{D}}$.

Since $\mathcal{A}_1$ and $\mathcal{A}_2$ have been 
computed in the derivative expansion of the 2+1d scalar QED
effective action \cite{cangemi1995derivative, gusynin1999derivative}, 
we simply quote the following result:
\begin{align}
\frac{e\tilde{\mathcal{A}_1}}{2}+\tilde{\mathcal{A}_2}=& 
e^2 \frac{i\xi}{8}\left[6\xi Y^3 -6Y^2 +4\xi Y-2 \right] \nonumber\\
=&\frac{e^2 w}{8} (R_3+R_1), 
\label{A1A2=}
\end{align}
where the simple form given in the last line is also known.
Here, $R_3$ and $R_1$ are the polynomials in the variables $\xi=-iw=eBT$ and $Y=\cot \xi$ defined in \eqref{Rp}.
Since the other terms $\tilde{\mathcal{B}}$, $\tilde{\mathcal{C}}$, and $\tilde{\mathcal{D}}$
can be also evaluated in simple forms in $R_2$ and $R_1$,
the polynomial forms $R_p$ and 
associated values after acting the linear operator $\mathcal{I}_{s'}(R_p)$ in \eqref{Is'}
are listed in TABLE \ref{tab:1}. 
The definition \eqref{Rp} of  $R_p$ is chosen such that the proper-time integral 
$\mathcal{I}_{s'}(R_p)$ can be evaluated systematically by integrations by parts.
 
Below, we use the propagator $g_{k\ell}(s_1, s_2)$ in \eqref{xixj} to evaluate 
$\mathcal{B}$, $\mathcal{C}$, and $\mathcal{D}$.
In the case of the pure magnetic field $F=B\varepsilon_{ij}$,  the index structure \eqref{gij} can be isolated as 
\begin{align}
\frac{g_{k\ell}(s_1,s_2)}{\tilde{g}(s_1,s_2)}&=\left(\delta_{k\ell}\cos eBs_{-} - \varepsilon_{k\ell}\sin eBs_{-}\right),
\label{gkl}
\end{align}
with $s_-=s_1-s_2$ and a scalar ${\tilde{g}(s_1,s_2)}$ given by
\begin{align}
eB{s_\xi}\ \tilde{g}(s_1,s_2)
&=\cos (\xi-|u_{-}|)   -     \cos (\xi-u_{+}),\nonumber\\
&\equiv \mathcal{G}(u_1,u_2),
\label{mathcalG}
\end{align}
where the propagator amplitude $\mathcal{G}(u_1,u_2)$, along with the scaled proper-time 
$u_n=eBs_n$ $(n=1,2)$ and $u_{\pm}=u_1\pm u_2$, is introduced.
Hereafter, we use the symbols $s_\xi=\sin \xi$ and $c_\xi=\cos \xi$.

The remaining contribution $\mathcal{B}$ in \eqref{B}, $\mathcal{D}$ in \eqref{D}, and $\mathcal{C}$ in \eqref{C} 
can now be evaluated via $L_{10}$ and $L_{11}$ in \eqref{L10}-\eqref{L11}
keeping the derivatives up to the second order, 
and then using the propagator \eqref{xixj} with its explicit form \eqref{gkl}-\eqref{mathcalG}.
The linear order term \eqref{B} yields,
\begin{align}
\mathcal{B} 
=&i\left(-\frac{\kappa e}{2}\right) \int_{0}^Tds\
\langle x_k(s) x_\ell(s) \rangle \partial_k \partial_\ell B   \nonumber\\
=&\frac{-i\kappa e}{2}\int_{0}^Tds~ ig_{k\ell}(s,s) \partial_k \partial_\ell B\nonumber\\
=&\frac{\kappa e}{2}\frac{\partial^2 B}{(eB)^2 s_\xi} \int_0^\xi du\ \mathcal{G}(u,u)\nonumber\\
=&\frac{\kappa e}{2}\frac{\partial^2 B}{(eB)^2 } \left[\xi Y - 1 \right],
\end{align}
where the computation is simplified since $s_{-}=u_{-}=0$.
By amputating the factor ${(\partial^2 B)}/{(eB)^2}$ in \eqref{A1B}, one obtains the corresponding tilde coefficient,
\begin{align}
\tilde{\mathcal{B}}= -\frac{\kappa e}{2}R_1, 
\label{B=}
\end{align}
where we used $R_1=-\xi Y +1$ from Table \ref{tab:1}.

Next, the diagonal term \eqref{D} yields,
\begin{align}
\mathcal{D}
=&
\frac{-(-\kappa e)^2}{2}\int_{0}^T\int_{0}^Tds_1ds_2\
\langle x_k(s_1) x_\ell(s_2) \rangle (\partial_k B)(\partial_\ell B)   \nonumber\\
=&\frac{-(\kappa e)^2}{2}  \int_{0}^T\int_{0}^Tds_1ds_2~ ig_{k\ell}(s_1,s_2) \partial_k B   \partial_\ell B \nonumber\\
=&\frac{(\partial B)^2}{(eB)^3 }\frac{-i(\kappa e)^2}{2s_\xi} \int_{0}^\xi\int_{0}^\xi dx_1dx_2\ \mathcal{G} (12) \cos u_{12},
\end{align}
where the notation $\mathcal{G}(12)=\mathcal{G}(u_1,u_2)$ is used.
By the amputation suggested in \eqref{A1B}, one obtains,
\begin{align}
\tilde{\mathcal{D}}
=&\frac{\kappa^2 e^2}{4} (i\xi)\left[ -\xi Y +1\right] \nonumber\\
=& \frac{\kappa^2 e^2}{4} w R_1.
\label{D=}
\end{align}

Similarly, the cross term \eqref{C} yields,
\begin{align}
\mathcal{C}=\frac{\kappa e^2}{3}\int_{0}^{T}\int_{0}^{T} ds_1ds_2\ & 
\langle x_i(s_1) \dot{x}_j(s_1)x_k(s_1) x_{\ell}(s_2)\rangle \nonumber\\
&\times \varepsilon_{ij}(\partial_k B) (\partial_{\ell} B).
\end{align}
To make the contractions with the velocity field $\dot{x}(s)$ along the magnon loop, one may use, for instance,
$\langle \dot{x}_j(s_1)x_{\ell}(s_2)\rangle=i \frac{d}{ds_1}g_{j\ell}(s_1,s_2)$.
By using \eqref{gkl}-\eqref{mathcalG} with the scale transform $s_n \to u_n$ 
and performing the same amputation of the factor ${(\partial B)^2}/{(eB)^3}$ 
as in $\mathcal{D} \to \tilde{\mathcal{D}}$, one obtains,
\begin{align} 
\tilde{\mathcal{C}}
&=\frac{-\kappa e^2}{3s^2_\xi}\int_0^\xi\int_0^\xi du_1du_2\ 
\Big( 4 \mathcal{G}(11)\mathcal{G}(12) \cos u_{-}   \nonumber\\
& \qquad\qquad      
 + \left[ \mathcal{G}(11)\mathcal{G}(\dot{1} 2 ) - \mathcal{G}(\dot{1}1)\mathcal{G}(12) \right] 
\sin u_{-} \Big)\nonumber\\
&=\frac{-\kappa e^2}{3s^2_\xi}  \left[ 3\xi^2 c^2_\xi -\frac{3\xi}{2}c_\xi s_\xi +\left(\frac{3\xi^2}{2}-\frac{3}{2}\right)s^2_\xi\right] \nonumber\\
&=  \frac{\kappa e^2}{2}  ( R_1 - w R_2),
\label{C=}
\end{align}
where in the first line the notations
$\mathcal{G}(\dot{1}2)=\frac{d}{du_1}\mathcal{G}(u_1,u_2)$ and
$\mathcal{G}(\dot{1}1)=\lim_{u_0\to u_1} \frac{d}{du_0}\mathcal{G}(u_0,u_1)$
are used.

For convenience, we collect the results \eqref{A1A2=}, \eqref{B=}, \eqref{C=}, and \eqref{D=} below:
\begin{align}
\frac{e\tilde{\mathcal{A}}_1}{2}+\tilde{\mathcal{A}}_2= \frac{e^2 w}{8} (R_3+R_1), \quad
\frac{e\tilde{\mathcal{B}}}{2} = -\frac{\kappa e^2}{4}R_1, \nonumber\\
\tilde{\mathcal{C}}= \frac{\kappa e^2}{2}  ( R_1 - w R_2), \quad
\tilde{\mathcal{D}}= \frac{\kappa^2 e^2}{4}w R_1
\label{collection}
\end{align}
Substituting this result \eqref{collection} in \eqref{A1B}
completes the evaluation of the combination
$\mathcal{S}=\frac{e}{2}\mathcal{C}_2+\mathcal{C}_3$ as given in \eqref{S=}.
 
We are now able to evaluate the coefficient ${\beta}$ for the derivative term in \eqref{V=i}
using the explicit form of the argument \eqref{S=} of the linear operator $\mathcal{I}_{s'=-\frac{1}{2}}(\mathcal{S})$
in \eqref{beta=sqrt}. By the integral transform definition of the operator \eqref{Is'}, it satisfies
$\mathcal{I}_{s'+1}(R_p)=\mathcal{I}_{s'}(wR_p)$.
We first use this property to eliminate $w$ and obtain,
\begin{align}
\sqrt{\frac{2}{\pi}} \beta =&\mathcal{I}_{-\frac{1}{2}}\!\! \left( 
\frac{w(R_3+R_1)}{8}  \!+\! \frac{\kappa({R_1}- 2w R_2)}{4}  \!+\! \frac{\kappa^2 }{4} w R_1
\right)\nonumber\\
=&\mathcal{I}_{+\frac{1}{2}}\!\left(\frac{R_3+R_1}{8}  
- \frac{\kappa}{2}  R_2 + \frac{\kappa^2 R_1}{4}\right)
\!+\frac{\kappa}{4} \mathcal{I}_{-\frac{1}{2}}\left( R_1\right) \nonumber\\
=&\frac{1}{8}\left( 20\zeta_{\frac{-3}{2}} - 18\zeta_{\frac{-1}{2}}+ 4\zeta_{\frac{1}{2}}  {~+~ } \zeta_{\frac{3}{2}}\right)\nonumber\\
&-\frac{\kappa}{2}\left( -6\zeta_{\frac{-1}{2}} + 
2\zeta_{\frac{1}{2}} + \frac{\zeta_{\frac{3}{2}} }{2}\right) \nonumber\\
&+\!\frac{\kappa^2}{4}\!\left(\!\zeta_{\frac{1}{2}} + \frac{\zeta_{\frac{3}{2}}}{2}\! \right)+
\frac{\kappa}{4}\!\left(\! -12\zeta_{\frac{-1}{2}} + 2\zeta_{\frac{1}{2}}\right),
\end{align}
where the shorthand notation $\zeta_x\equiv \zeta(x)$ is used.
In the second line, we used the analytically continued values listed in TABLE \ref{tab:1}.

For the quantum soliton physics, we may use $\kappa=1$.
This further simplifies the above expression, and we obtain the value as shown in \eqref{beta=zeta}:
\begin{align}
{\beta}=&
 \sqrt{\frac{\pi}{2}}\frac{10\zeta\left(\frac{-3}{2}\right)- 9\zeta\left(\frac{-1}{2}\right) + \zeta\left(\frac{1}{2}\right)  }{4}
\nonumber\\
=&\frac{8 \pi ^2 \zeta \left(\frac{1}{2}\right)+18 \pi  \zeta \left(\frac{3}{2}\right)-15 \zeta \left(\frac{5}{2}\right)}{32 \sqrt{2} \pi ^{3/2}},
\end{align}
where the functional equation for the zeta function 
$\zeta(1-s)=2(2\pi)^{-s}\cos(\frac{\pi s}{2})\Gamma(s)\zeta(s)$ is used for turning $\zeta(s)$'s in the first line
into those with the positive arguments.

\bibliography{_MagnonReferencesFinal}


\end{document}